\documentclass[12pt]{article}
\newcommand{\drawsquare}[2]{\hbox{%
\rule{#2pt}{#1pt}\hskip-#2pt
\rule{#1pt}{#2pt}\hskip-#1pt
\rule[#1pt]{#1pt}{#2pt}}\rule[#1pt]{#2pt}{#2pt}\hskip-#2pt
\rule{#2pt}{#1pt}}

\newcommand{\Yfund}{\raisebox{-.5pt}{\drawsquare{6.5}{0.4}}}
\newcommand{\Ysymm}{\raisebox{-.5pt}{\drawsquare{6.5}{0.4}}\hskip-0.4pt%
        \raisebox{-.5pt}{\drawsquare{6.5}{0.4}}}
\newcommand{\Ythrees}{\raisebox{-.5pt}{\drawsquare{6.5}{0.4}}\hskip-0.4pt%
          \raisebox{-.5pt}{\drawsquare{6.5}{0.4}}\hskip-0.4pt%
          \raisebox{-.5pt}{\drawsquare{6.5}{0.4}}}
\newcommand{\Yasymm}{\raisebox{-3.5pt}{\drawsquare{6.5}{0.4}}\hskip-6.9pt%
        \raisebox{3pt}{\drawsquare{6.5}{0.4}}}
\newcommand{\Ythreea}{\raisebox{-3.5pt}{\drawsquare{6.5}{0.4}}\hskip-6.9pt%
        \raisebox{3pt}{\drawsquare{6.5}{0.4}}\hskip-6.9pt
        \raisebox{9.5pt}{\drawsquare{6.5}{0.4}}}
\newcommand{\Yfoura}{\raisebox{-3.5pt}{\drawsquare{6.5}{0.4}}\hskip-6.9pt%
        \raisebox{3pt}{\drawsquare{6.5}{0.4}}\hskip-6.9pt
        \raisebox{9.5pt}{\drawsquare{6.5}{0.4}}\hskip-6.9pt
        \raisebox{16pt}{\drawsquare{6.5}{0.4}}}
\newcommand{\Ysquare}{\raisebox{-3.5pt}{\drawsquare{6.5}{0.4}}\hskip-0.4pt%
        \raisebox{-3.5pt}{\drawsquare{6.5}{0.4}}\hskip-13.4pt%
        \raisebox{3pt}{\drawsquare{6.5}{0.4}}\hskip-0.4pt%
        \raisebox{3pt}{\drawsquare{6.5}{0.4}}}
\newcommand{\Yoneoone}{\raisebox{-3.5pt}{\drawsquare{6.5}{0.4}}\hskip-6.9pt%
        \raisebox{3pt}{\drawsquare{6.5}{0.4}}\hskip-6.9pt%
        \raisebox{9.5pt}{\drawsquare{6.5}{0.4}}\hskip-0.4pt%
        \raisebox{9.5pt}{\drawsquare{6.5}{0.4}}}%
\newcommand{\Ytwoone}{\raisebox{-3.5pt}{\drawsquare{6.5}{0.4}}\hskip-6.9pt%
        \raisebox{3pt}{\drawsquare{6.5}{0.4}}\hskip-0.4pt%
        \raisebox{3pt}{\drawsquare{6.5}{0.4}}\hskip-0.4pt%
        \raisebox{3pt}{\drawsquare{6.5}{0.4}}}%

\newcommand{\jref}[4]{{\it #1} {\bf #2}, #3 (#4)}

\newcommand{\MPLA}[3]{\jref{Mod.\ Phys.\ Lett.}{A#1}{#2}{#3}}
\newcommand{\NC}[3]{\jref{Nuovo Cim.}{#1}{#2}{#3}}
\newcommand{\NPB}[3]{\jref{Nucl.\ Phys.}{B#1}{#2}{#3}}

\newcommand{\PLB}[3]{\jref{Phys.\ Lett.}{#1B}{#2}{#3}}

\newcommand{\PRD}[3]{\jref{Phys.\ Rev.}{D#1}{#2}{#3}}

\newcommand{\PRL}[3]{\jref{Phys.\ Rev.\ Lett.}{#1}{#2}{#3}}

\renewcommand{\theequation}{\thesection.\arabic{equation}}
\makeatletter
\def\vereq#1#2{\lower3pt\vbox{\baselineskip1.5pt \lineskip1.5pt
\ialign{$\m@th#1\hfill##\hfil$\crcr#2\crcr\sim\crcr}}}
\makeatother

\begin{document}

\begin{titlepage}
\begin{center}
\today     \hfill    LBNL-40898 \\
~{} \hfill UCB-PTH-97/52  \\
~{} \hfill hep-th/9710105\\

\vskip .3in

{\Large \bf Discrete Anomaly Matching\footnote{This work was
  supported in part by the U.S. 
Department of Energy under Contracts DE-AC03-76SF00098, and in part by the 
National Science Foundation under grant PHY-95-14797.}}

\vskip 0.3in

{\bf Csaba Cs\'aki\footnote{Research fellow, Miller Institute for 
Basic Research in Science.} and Hitoshi Murayama\footnote{Supported in 
part by an Alfred P. Sloan Foundation Fellowship.}}

\vskip 0.15in

{\em Theoretical Physics Group\\
     Ernest Orlando Lawrence Berkeley National Laboratory\\
     University of California, Berkeley, California 94720}

\vskip 0.1in
{\rm and}
\vskip 0.1in

{\em Department of Physics\\
     University of California, Berkeley, California 94720}

\vskip 0.1in
{\tt  csaki@thwk5.lbl.gov, murayama@thsrv.lbl.gov}

\end{center}

\vskip .25in

\begin{abstract}
We extend the well-known 't Hooft anomaly matching conditions for
continuous global symmetries to discrete groups.
We state the matching conditions for all possible anomalies which 
involve discrete symmetries explicitly in Table~\ref{tab:summary}.
There are two types of discrete anomalies.  For Type I anomalies, 
the matching conditions have to be always satisfied regardless of
the details of the massive bound state spectrum. The Type II anomalies
have to be also matched except if there are fractionally charged massive
bound states in the theory.  We check discrete anomaly
matching in recent solutions of certain 
$N=1$ supersymmetric gauge theories, most of which satisfy these
constraints. The excluded examples include the chirally symmetric phase
of $N=1$ pure supersymmetric Yang-Mills theories described by the
Veneziano--Yankielowicz Lagrangian and certain
non-supersymmetric confining theories.  The conjectured self-dual
theories based on exceptional gauge groups do not satisfy discrete 
anomaly matching nor mapping of operators, 
and are viable only if the 
discrete symmetry in the electric theory appears as an accidental 
symmetry in the magnetic theory and vice versa.
\end{abstract}

\end{titlepage}

\newpage

\section{Introduction}
\setcounter{equation}{0}
\setcounter{footnote}{0}

Understanding the dynamics of any physical system beyond perturbation
theory has always been a challenging task.  
In the context of quantum
field theories, many techniques have been developed to attack the
problem with various degrees of success and applicability: computer
simulation of field theories on a lattice, exact solutions using Bethe
Ansatz or Yang-Baxter equation, mean field approximation,
Schwinger--Dyson equation, large $N$, etc.  If the system possesses a 
relatively
large global symmetry, however, one of the most powerful method to
study a possible low-energy spectrum of a theory is 't Hooft anomaly
matching~\cite{tHooft}.  
This method has been especially useful in the recent
remarkable progress in supersymmetric gauge
theories~\cite{Seiberg,IntrSeib,IntrPoul,Kutasov,otherKutasov,
G2,SOduals,deconf,SUantisymm,sconf,DM,GN}.

In the method of 't Hooft anomaly matching, one compares the anomalies
of the global symmetries in a model between the fundamental theory and a
proposed low-energy theory.  If the symmetries are not spontaneously
broken, it is argued that the anomalies must match.  Since the
matching of anomalies often involve linear and cubic equations, the
requirement that the anomalies must match usually results in a highly
non-trivial consistency check of a candidate low-energy theory.  Being
a necessary condition, anomaly matching can not establish that the
candidate theory is indeed the correct low-energy description of the
given fundamental theory; it can however either be used to exclude a
proposed candidate or to give a strong support for it.  Because of this
nature of the method, it is very important to exploit all possible
available constraints.  Even when only one of the constraints fails, it
excludes the proposed low-energy theory.

In this paper, we will 
show that discrete symmetries also have anomalies which have
to be matched between the fundamental and low-energy
theories.\footnote{In this paper, we focus only on Abelian discrete
symmetries while we believe that our arguments for discrete anomaly
matching could be extended to non-Abelian discrete symmetries as well.}  The
argument can be summarized as follows.  A discrete symmetry can be
promoted to a continuous  global symmetry by regarding certain
couplings of the theory as background fields. This new continuous
global symmetry has to satisfy the usual 't Hooft anomaly matching
conditions. Once the background field is frozen to its actual value,
the continuous symmetry is broken to a
discrete one.  However, the anomaly matching conditions must still be
satisfied mod $N$ for a $Z_N$ symmetry, if one uses a normalization
where all $Z_N$ charges are integers. 
Furthermore, 
one can work out how the decoupling of massive fields can modify the
anomalies and the possible modifications
can be classified.  The combination of the original anomaly matching
and the decoupling of heavy fields give powerful constraints on the
low-energy particle content.

We apply the matching of discrete anomalies to many models studied in
the literature.  We find that all Seiberg dualities in $N=1$
supersymmetric gauge theories~\cite{Seiberg,IntrSeib,IntrPoul} 
match discrete anomalies in a highly
non-trivial manner.  We also find that certain conjectured
dynamics of gauge theories can be excluded by this consideration.  The
examples include chirally symmetric vacua in $N=1$ supersymmetric
Yang--Mills theories \cite{Shifman} described by the 
Veneziano--Yankielowicz Lagrangian \cite{VY}, certain 
non-supersymmetric chiral gauge 
theories \cite{Albright,othernonsusy}
and self-dual supersymmetric theories based on exceptional groups 
\cite{Ramond,Distler,Karch}.

The paper is organized as follows. In Section~\ref{sec:tHooft}, we review 't
Hooft anomaly matching for continuous global symmetries. In
Section~\ref{sec:discrete}, we discuss the possible kinds of discrete
symmetries. In Section~\ref{sec:matching}, we give our arguments for discrete
anomaly matching and discuss the decoupling of heavy fermions. The
final form of the discrete anomaly matching conditions are given in  an
explicit form at the end of Section~\ref{sec:recap}. In 
Section~\ref{sec:SO}, we
apply the discrete anomaly matching conditions to the
Intriligator--Seiberg solution~\cite{IntrSeib} of $N=1$ supersymmetric
$SO(N)$ theories with vectors. In Section~\ref{sec:other}, other supersymmetric
examples are discussed. Section~\ref{sec:exclude} 
gives the examples excluded by
the discrete anomaly matching conditions. Finally, we conclude in
Section~\ref{sec:concl}.

\section{'t Hooft Anomaly Matching\label{sec:tHooft}}
\setcounter{equation}{0}
\setcounter{footnote}{0}

One of the most powerful tools for studying the 
non-perturbative
low-energy dynamics of strongly interacting gauge theories are the 
't Hooft anomaly matching conditions. These are highly non-trivial
constraints on the massless fermion content of a confining theory,
and it can also be used as a check for conjectured dualities.
Since the 't Hooft anomaly matching conditions play a central role in
our discussion, 
in this section  we briefly summarize 't Hooft's 
original argument~\cite{tHooft} about matching of continuous global anomalies.

Let us assume that we have a strongly interacting gauge theory, with
gauge group $G_{gauge}$, and that there is a continuous non-anomalous
global symmetry $G_{global}$. Since $G_{global}$ is a global symmetry,
there is generally no reason for the $G_{global}^3$ anomaly to be 
vanishing. One can imagine however to include spectator fields
which transform under $G_{global}$ but not under $G_{gauge}$,
such that the  $G_{global}^3$ anomaly is exactly canceled. Then one
can weakly gauge the $G_{global}$ group as well. Let us now consider the
low-energy effective theory which contains some massless fermions, which
are to be thought of as composites of the original degrees of freedom.
Since $G_{global}$ is weakly gauged, it has to be anomaly free in the
low-energy effective theory as well. However, since the spectator 
fields do not transform under $G_{gauge}$, they do not participate in the
strong dynamics, and hence their contribution to the $G_{global}^3$ anomaly 
is identical in the high-energy and the low-energy descriptions.
Therefore the $G_{global}^3$ anomaly of the original degrees of freedom
(excluding the spectators) must exactly match the contribution of the
composite fields in the low-energy theory. This argument can be trivially
generalized to show that, in the case the global symmetry is a product
group of the form $G_{global}=G_1\times G_2 \times \ldots \times U(1)_1\times 
U(1)_2 \times \ldots$, all the $G_i^3$, $G_i^2U(1)_j$, $U(1)_iU(1)_jU(1)_k$
as well as the $U(1) ({\rm gravity})^2$ anomalies must match between the
high-energy and the low-energy theories.  Here and below, $G_i$ refer
to simple groups.  
To be more explicit, we list below the quantities whose values calculated
in the high-energy and low-energy theories have to be precisely equal:
\begin{eqnarray}
\label{matching}
& G_i^3: & \sum_R A^i_R \nonumber \\
& G_i^2U(1)_j: & \sum_R \mu_R^iq_R^j \nonumber \\
& U(1)_iU(1)_jU(1)_k: & \sum_R q_R^iq_R^jq_R^k \nonumber \\
& U(1)({\rm gravity})^2: & \sum_R q_R^i 
\end{eqnarray}
where $A$ is the cubic anomaly coefficient defined by the relation
${\rm Tr}_R\, \{ T^a,T^b\} T^c=A_Rd^{abc}$ (the
$T$'s being the generators of the group $G_i$ in a given representation $R$),
$\mu_R$ is the Dynkin index ${\rm Tr}_R\, T^aT^b=\mu_R \delta^{ab}$,  
and $q^i$'s are the
$U(1)_i$ charges. The sum over $R$ denotes the summation over all 
representations
of fermions present in the high-energy or the low-energy descriptions.

The fact that these constraints are
satisfied is the most important evidence in favor of the
low-energy solutions of certain $N=1$ supersymmetric gauge theories
proposed by Seiberg and others. However, many of these theories have
discrete global symmetries in addition to the continuous ones. It is a
natural question to ask whether the presence of the discrete
symmetries further constrains the low-energy spectrum. We will
next show that this is indeed the case: discrete symmetries
have to obey certain anomaly matching conditions as well. In the next
section we first review the different types of discrete symmetries a
theory can have and their possible origins. Then in Section~\ref{sec:matching}
we show what the anomaly matching conditions for discrete symmetries are.

\section{Discrete Symmetries\label{sec:discrete}}
\setcounter{equation}{0}
\setcounter{footnote}{0}

In this section, we review the possible origins of discrete symmetries
in a quantum field theory. This is useful in order to find all
non-trivial discrete symmetries of a given theory. There are
two different types of discrete symmetries. One type is when 
the discrete symmetry commutes with the gauge group. We call these
the ``flavor-type'' discrete symmetries. The other type is when the 
the discrete symmetries do not commute with the gauge transformations, 
and are given by outer automorphisms of the Lie algebras.  We call them ``color
conjugation'' type discrete symmetries.  At the end of the section we 
discuss when the discrete symmetries are independent from 
the center of the continuous global symmetries.

\subsection{Flavor-Type Discrete Symmetries}

Flavor-type discrete symmetries arise when a continuous flavor 
symmetry of the kinetic terms of the Lagrangian is broken explicitly 
or spontaneously, but a discrete subgroup of the continuous symmetry 
is left unbroken.  We review the possible mechanisms for breaking a 
continuous flavor symmetry to its discrete subgroup below.

\subsubsection{Explicit Breaking}

The simplest possibility is that a continuous global symmetry is 
explicitly broken by an interaction term in the Lagrangian. 
For example, if there is a global $U(1)$ symmetry, under which the
fields $\phi_i$ (which could be either a bosonic or a fermionic
field) have charge $q_i$, an interaction term
\begin{equation} 
{\cal L}_{\it break}=\prod_i \phi_i, \qquad \sum_iq_i\neq 0,
\end{equation}
breaks the global $U(1)$ to its $Z_N$ subgroup with
$N= \sum_iq_i$. The fields $\phi_i$ transform under this $Z_N$ as 
\begin{equation} 
        \phi_i \to e^{\frac{n q_i}{N}2\pi i}\phi_i,\qquad n=0,1,\ldots, N-1. 
\end{equation}

\subsubsection{Breaking by Instantons}
This happens when a global $U(1)$ is anomalous. Assume that 
the left-handed Weyl fermion fields $\psi_i$ carry charges 
$q_i$ of a classical $U(1)$ symmetry, 
and that this $U(1)$ is anomalous under the gauge group. One 
consequence of the anomaly is that the correlator 
\begin{equation}
\label{corr}
\langle \prod_i \psi_i^{\mu_i} \rangle 
\end{equation}
does not vanish in an instanton background~\cite{tHooft2}, 
thus breaking the
anomalous $U(1)$ symmetry. Here $\mu$ is the Dynkin index of the given
fermion under the gauge group, where the index is defined as 
${\rm Tr} T^aT^b=\mu \delta^{ab}$. The $T^a$'s are the generators of
the gauge group in the representation of the fermion $\psi_i$. The
normalization of the Dynkin index is chosen such that it exactly
corresponds to the number of fermion zero modes in a one-instanton
background ({\it i.e.}\/ the index of the 
fundamental of $SU$ and $Sp$ is normalized to one while the vector of
$SO$ is normalized to two). However, the correlator in 
Eq.~(\ref{corr})
is invariant under the discrete transformations 
\begin{equation} 
\psi_i\to e^{\frac{nq_i}{N}2\pi i}\psi_i, \qquad
n=0,1,\ldots ,N-1,
\end{equation}
thus a discrete $Z_N$ subgroup with  $N=\sum_i q_i \mu_i$
is left unbroken. 

\subsubsection{Spontaneous Breaking}

It is also possible that a continuous global symmetry is
spontaneously broken by an expectation value of one of the
fields, but that the VEV of the field leaves a discrete rotation
invariant. The general rule for the $U(1)\to Z_N$ type breaking is that
if the field $\varphi$ with non-vanishing expectation value 
$\langle \varphi \rangle \neq 0$ has charge $N$ under a
global $U(1)$ symmetry, then the $Z_N$ subgroup of $U(1)$ under which
\begin{equation} 
\psi_i \to e^{\frac{nq_i}{N}2\pi i} \psi_i,\; \; \; n=0,1,\ldots ,N-1
\end{equation} 
is left unbroken. Note that this transformation has no
effect on the field $\varphi$ with the non-vanishing expectation value
as required.

\subsection{Color Conjugation Type Discrete Symmetries\label{sec:conjug}}

The flavor-type discrete symmetries considered above all arise from 
breaking of the continuous flavor symmetries of the kinetic terms of 
the Lagrangian.  However, it is possible that a theory has more 
symmetries than the usual global flavor symmetries.  If such 
symmetries are present, they can not commute with the gauge group; 
otherwise they would be contained in the flavor symmetries of the 
theory.  In order for such transformations to be symmetries of the 
theory, they must leave the Lie algebra of the gauge group 
invariant, and hence they must be outer automorphisms of the Lie algebra.  
The complete list of all possible outer automorphisms of 
simple gauge groups is given by (see $e.g.$~\cite{auto}):
\begin{eqnarray}
  \begin{array}{cl}
        SU(N): & Z_2 \quad (N>2) \\
        SO(2N): & Z_2  \quad (N>2)  \\
        E_6: & Z_2   \\
        SO(8): & S_3 
  \end{array} 
\end{eqnarray}
while the other simple gauge groups do not have a non-trivial outer
automorphism.  We call discrete symmetries based on these outer
automorphisms ``color conjugation type'' discrete symmetries.

As the name suggests, a color conjugation is a generalization of the
familiar charge conjugation in QED.  Charge conjugation changes
the sign of the electric charge $Q \to - Q$, and interchanges charge
$+1$ fields with charge $-1$ fields.  An immediate generalization of
this for non-Abelian gauge groups is given by
\begin{equation}
\label{cc}
 T^a\to {\cal C}^{-1} T^{a} {\cal C} = -T^{a*},
\end{equation}
where the $T^a$'s are the generators of the gauge group and ${\cal C}$
is the charge conjugation operator.  
For the case of $SU(N)$, $E_6$ and $SO(4k+2)$ gauge groups, this
indeed defines outer automorphisms on the Lie algebras, and we call
this transformation ${\cal C}$ charge conjugation.  Charge 
conjugation exchanges representations with their complex conjugates.
Note that the charge conjugation is trivial for real representations,
and is equivalent to a gauge transformation for pseudo-real
representations.

There is, however, another way to generalize the charge conjugation in
QED to $SO(N)$ groups.  With fields $q^+$ and $q^-$ with 
electric charges $\pm 1$, one can define an $SO(2)$ doublet 
$\big(q^1, q^2\big) =
\big(i(q^+ -q^-), q^+ + q^-\big)/\sqrt{2}$, on which the charge
conjugation 
acts as
the sign flip of the first ``color'', $q^1 \to - q^1$, $q^2 \to q^2$.
In general, we define a ``color-parity'' transformation ${\cal P}$
on an $SO(N)$
vector by flipping the sign of one particular color (for example the
first color),
which defines an automorphism of the Lie algebra
\begin{equation}
\label{parity}
 M_{ij}\to {\cal P}^{-1} M_{ij} {\cal P} = \left\{
 \begin{array}{ll} -M_{1j} & (j\neq 1)\\
 M_{ij} & (i,j\neq 1)
 \end{array} \right. ,
\end{equation}
where the $M_{ij}$'s are the $SO(N)$ generators.
One can view the color-parity as a non-trivial 
element of the $O(N)$ extension of $SO(N)$ group, {\it i.e.}\/, a 
parity-like transformation.\footnote{For $SO(2N+1)$ groups, this 
parity-like transformation is gauge equivalent to an overall sign flip 
of the vector, and hence is of the flavor-type.}  
As we show in \ref{app:charge}, this
definition of the color-parity transformation 
is equivalent to the charge conjugation of Eq.~(\ref{cc}) for $SO(4k+2)$ 
gauge groups up to gauge transformations. 
On the other hand, $SO(4k)$ groups have only real or pseudo-real
representations, and hence the charge conjugation Eq.~(\ref{cc}) 
is equivalent to an $SO(4k)$ gauge transformation and thus is not an
outer automorphism of the Lie algebra.  The color-parity
transformation is an outer automorphism for all $SO(2N)$ gauge groups, and 
interchange two inequivalent spinor representations (often
referred to as spinor and conjugate-spinor representations). 

Therefore, it is convenient to define the color conjugations by the
charge conjugation (\ref{cc}) for $SU(N)$ and $E_6$, and by the
color-parity (\ref{parity}) for $SO(2N)$ groups.
The $SO(8)$ group is special and its
$S_3$ automorphism is the triality permuting the vector, spinor
and conjugate spinor representations. 

Note that color conjugation type discrete symmetries
are not necessarily realized within a
theory, but they may map the given theory to another one.
Whether this is the case depends
on the matter content of the theory. Since the color conjugations
usually interchange representations, 
only non-chiral theories have these extra
discrete symmetries; in chiral theories these symmetries are
broken by the matter content.  Even if a color conjugation
can be defined in a given theory, it is usually not very
useful from the point of view of anomaly matching, since the notion of
anomaly is hard to define if a symmetry interchanges
representations.  Thus the only interesting case is if one has a gauge
group with a non-trivial outer automorphism and only self-conjugate
representations. In this case, the color conjugation symmetries can mix
with the usual flavor type discrete symmetries and may be
important. We will see several examples of this happening in the
supersymmetric $SO(N)$ examples of Sections~\ref{sec:SO}.

\subsection{Independence of Discrete Symmetries\label{sec:indep}}

We have seen above how to find the discrete symmetries of a given theory.
However, one has to be careful with the identification of the
non-trivial discrete symmetries. The reason is that a discrete symmetry might 
be contained in a continuous symmetry as its discrete
subgroup. As an example, let us consider  QCD with $F$ flavors. 
The classical theory has the continuous symmetries
\begin{equation} 
\begin{array}{c|ccccc} 
& SU(N) & SU(F)_{Q} & SU(F)_{\bar{Q}} & U(1)_B & U(1)_A \\ \hline 
Q & \Yfund & \Yfund & 1 & 1 & 1 \\
\bar{Q} & \overline{\Yfund} & 1 & \Yfund & -1 & 1 \end{array}\qquad , 
\end{equation}
where $SU(N)$ is the gauge group, the two $SU(F)$ factors are 
the non-Abelian global symmetries which transform either 
the quarks or the antiquarks\footnote{Here
and throughout the paper, a fermion means a left-handed two-component
Weyl spinor.} 
among each other, $U(1)_B$ is the baryon number, and $U(1)_A$ is the 
axial $U(1)$ under which both quarks and antiquarks transform by the same phase.
This $U(1)_A$ is however anomalous, since the correlator
\begin{equation}
 \label{QCDanomaly}
 \langle Q^F \bar{Q}^F \rangle \neq 0 
\end{equation}
is non-vanishing. Thus  $U(1)_A$ is broken by instantons to its
$Z_{2F}$ discrete subgroup, under which 
\begin{equation} 
  Q\to e^{2\pi in/2F} Q, \quad \bar{Q}\to  
  e^{2\pi in/2F} \bar{Q}, \qquad n=0,1,\ldots ,2F-1. 
\end{equation}
However this $Z_{2F}$ symmetry is 
not a new symmetry of the theory. The reason is that one can choose
a discrete subgroup of the continuous flavor symmetries which
exactly coincides with this $Z_{2F}$ symmetry. Take for example the
center of one of the $SU(F)_{Q}$ flavor symmetries, which is a 
$Z_F$ transformation under which only the quarks $Q$ transform with
charge one. A combination of this with discrete baryon number transformation
with a phase  $-\pi /F$ is exactly the $Z_{2F}$ symmetry
from the anomalous $U(1)$, thus it is part of the other continuous 
global symmetries.  In the absence of explicit breaking terms,
only those theories which do not contain matter fields in the fundamental 
representations (Dynkin index one) have non-trivial discrete
symmetries from anomalous $U(1)$'s.

The complete list of the centers of simple groups is given in
\ref{app:centers}.

\section{Discrete Anomaly Matching\label{sec:matching}}
\setcounter{equation}{0}
\setcounter{footnote}{0}

In this section, we will show that discrete global symmetries have to obey
anomaly matching constraints as well. We will give two different
arguments for this. One argument is based on considering correlators
in instanton backgrounds after gauging a non-Abelian flavor symmetry 
and is the natural generalization of 't Hooft's original argument 
summarized in Section~\ref{sec:tHooft}. In the second argument, we promote the
coupling which breaks the continuous global symmetry to its discrete
subgroup to a background field, thus restoring the full continuous
global symmetry. 

We have to note that anomalies of discrete symmetries 
have been considered previously 
in Refs.~\cite{PTWW,IR,BD}. In these papers, the authors considered
the consequences of ``gauging'' the discrete symmetries \cite{KW}
on the low-energy
spectrum. The assumption was that all global symmetries are broken by
quantum gravitational effects, and hence only ``gauged'' discrete symmetries
can be realized on a realistic low-energy particle
spectrum. They then considered the consequences of the anomaly
cancellation for the gauged discrete symmetries.  For our purpose,
we do not assume
that discrete symmetries have to be anomaly free. Instead, we are going
to compare the anomalies of the discrete symmetries between the high-energy 
and the low-energy theories. Even though the spirit of our work
is very different from that of Refs.~\cite{PTWW,IR,BD}, the arguments
below for discrete anomaly matching will be in many aspects similar to
those in~\cite{PTWW,IR,BD}.   
Our first argument for discrete anomaly matching is based on
instantons and resembles the 
spirit of Refs.~\cite{PTWW,BD}, while our second
argument of restoring the continuous symmetry is closer to the
attitude of Ref.~\cite{IR}.

We will present two classes of anomaly matching constraints. The Type
I constraints (which include the $G_F^2Z_N$ and $Z_N({\rm gravity})^2$
anomalies) have to be satisfied independently of any assumptions about
the massive particle spectrum. The Type II constraints (which include
the $Z_N^3$, $U(1)^2Z_N$, $U(1)Z_N^2$, $Z_N^2Z_M$ and $Z_NZ_MU(1)$
anomalies), however, may be evaded if the massive spectrum contains
fractionally charged particles. We will discuss the issue of
charge fractionalization in more detail in Section~\ref{fractional}.

\subsection{The Instanton Argument\label{sec:inst}}

Here we will show that the discrete $G_F^2Z_N$ and $Z_N({\rm gravity})^{2}$
anomalies have to be matched between the high-energy
and the low-energy theories (Type I constraints), where $G_F$ is a 
non-Abelian flavor
symmetry. Since the $Z_N$ charges of the fields are defined only mod
$N$, the most stringent constraint we can expect is anomaly
matching mod $N$. We will see that for the $G_F^2Z_N$ anomaly this is
indeed the case, while for $Z_N({\rm gravity})^{2}$ 
the anomalies have to match only mod $N/2$, if $N$ is
even (and mod $N$ if $N$ is odd). 

Let us first discuss the $G_F^2Z_N$ anomaly. We assume that the theory
we consider has a $G_F\times Z_N$ discrete symmetry which is
non-anomalous under the gauge group. In general, the $G_F^3$ and the
$G_F^2Z_N$ anomalies do not necessarily vanish. Next we introduce
spectator fields which do not transform under the gauge group such that
both the $G_F^3$ and the $G_F^2Z_N$ anomalies vanish (the latter mod
$N$). Then we can weakly gauge the $G_F$ group since the $G_F^3$ anomaly
now vanishes.  Because the $G_F^2Z_N$ anomaly vanishes as well,
this means that the $Z_N$ symmetry is unbroken in a background of
$G_F$ instantons; it is an exact symmetry of the theory. 
In other words, the non-vanishing correlator in a $G_F$ 
instanton background $\langle \prod_i \psi_i^{\mu_i} \rangle $ is 
invariant under $Z_N$.  Thus $\sum_i \mu_i q_i=0$ mod $N$, where
$\mu_i$ is the Dynkin index of the Weyl fermions under $G_F$, while
$q_i$'s are the $Z_N$ charges. 

If the $Z_N$ symmetry is not spontaneously broken, then the low-energy
effective theory must have $Z_N$ as an unbroken exact symmetry as well. This
means that the non-vanishing correlator in the $G_F$
instanton background calculated for the low-energy bound states must
also be invariant under $Z_N$. Thus we conclude that $\sum_i \mu_i
q_i=0$ mod $N$ in the low-energy theory as well. Since the spectators
do not transform under the gauge group, they do not participate in
forming the bound states, and their contribution to the $G_F^2Z_N$ 
discrete anomaly is the same in the high-energy and in the low-energy
theories. Therefore, the bound states must match the $G_F^2Z_N$
anomaly of the original degrees of freedom mod $N$. 

One can repeat exactly the same argument for the $Z_N({\rm gravity})^{2}$ 
anomaly (which will constrain the $\sum_i q_i$
quantity, where $q_i$ are the $Z_N$ charges) by considering
correlators in gravitational instanton backgrounds. 
One has to be, however, careful with identifying the correct anomaly
matching condition, because there are always even number of zero
modes for a Weyl fermion in a gravitational instanton background. This
is due to Rohlin's theorem~\cite{Rohlin} which states that the signature of a
smooth, compact, spin four-manifold is divisible by 16. Since the
$\hat{A}$-genus of a four-dimensional manifold is an eighth of
the signature (see, {\it e.g.}\/ \cite{EGH}), there are always even
numbers of zero modes for a Weyl fermion. 
The smallest number of zero modes is found, for instance, 
on a $K3$ manifold, which give two zero modes for every Weyl fermion.
Even if the
$Z_N({\rm gravity})^{2}$ anomalies differ by $N/2$ between
fundamental and low-energy theories, it does not change the conclusion that the 
$Z_N$ symmetry is not broken either in the high-energy or in the
low-energy theory by gravitational instantons. Thus the 
$Z_N({\rm gravity})^{2}$ anomaly has to be matched only mod
$N/2$, if $N$ is even. 

The origin of the possible difference of $N/2$ in the 
$Z_N({\rm gravity})^{2}$ anomaly can also be understood by
considering decoupling of heavy fermions~\cite{IR}. 
The contribution of such particles to 
continuous anomalies is always vanishing. This is not the case for discrete
anomalies and the possible contributions of such particles must be 
enumerated. One can have, for example, a pair of different Weyl fermions pairing 
up and getting a Dirac mass.
In this case, the charges of these fermions
must obey $q_1+q_2=mN$, where $q_1,q_2$ and $m$ are integers. 
Then these
fermions contribute integer multiples of $N$ to all anomalies, and since all
the anomaly matching equations are modulo $N$ anyway, such particles do
not change these equations. However for even $N$, there is another 
possibility: 
a single 
fermion with $Z_N$ charge $\frac{m}{2}N$ can acquire a Majorana mass. 
In this case,
the contribution of this fermion to the 
$Z_N({\rm gravity})^{2}$ anomaly is $\frac{m}{2}N$,
thus there can be a difference which is a half-integer multiple
of $N$ between the high-energy and the low-energy values of the
$Z_N({\rm gravity})^{2}$ anomaly. 
The possible existence of such massive Majorana fermions leads to the
weaker anomaly matching condition for the $Z_N({\rm gravity})^{2}$ 
anomaly. On the other hand, this also means that we might
gain some information (even if very limited) about the massive
spectrum as well. If the anomalies match only mod $N/2$, then we can
conclude that there must be odd number of massive Majorana fermions
with $Z_N$ charge $N/2$ present in the theory. Such Majorana fermions
however do not weaken the $G_F^2Z_N$ anomaly matching constraint,
since the Dynkin indices of real representations (as required for a 
Majorana fermion) are even, and
therefore the contribution of heavy Majorana particles to the $G_F^2Z_N$
anomalies is a multiple of $N$. 

\subsection{The Spurion Argument}

We have shown above that the discrete $G_F^2Z_N$ and 
$Z_N({\rm gravity})^{2}$ anomalies have to be matched mod $N$
and mod $N/2$ between the low-energy and the high-energy theories. 
We used the fact that we can study correlators in the $G_{F}$ or gravitational
instanton background. However, this argument can
obviously not be extended to the Type II anomalies, such as $U(1)^2Z_N$, 
$Z_N^3$. Therefore we present another argument, which will show
that Type II discrete anomalies have to be matched as well, not just the two
discussed in the previous section, assuming there are no massive
states with fractional charges. Note that the matching of the Type I
anomalies is independent of the details of the massive spectrum. 

We discuss only flavor-type discrete symmetries, which arise 
due to the breaking  of a continuous global symmetry via an
interaction term in the Lagrangian.\footnote{It is straightforward to 
generalize the discussion to the color conjugation type discrete 
symmetries by enlarging the gauge group but breaking it by a spurion.}
This continuous global symmetry can
be restored, if we promote the coupling which breaks the continuous
symmetry to a background field (``spurion''). 
For example in the case of explicit breaking by the interaction
\begin{equation} 
  {\cal L}_{\it break}=\lambda \prod_i \phi_i, \qquad \sum_iq_i\neq 0,
\end{equation}
we can assign $U(1)$ charge $-\sum_iq_i$ to the coupling $\lambda$.

In the case of a discrete $Z_N$ symmetry arising from an anomalous
$U(1)$ symmetry, we can first add a pair of fermions $\psi_0$ and 
$\psi_{-N}$ which exactly
cancel the $U(1)G_{gauge}^2$ anomaly.
For example for $SU(n)$ we add a fundamental with charge $0$ and an
antifundamental with charge $-N$, for $Sp(2n)$ we add a fundamental
with charge $0$ and another one with charge $-N$, and for $SO(n)$ we add one
vector with charge $-N/2$. This latter is
allowed because $N$ is even in the case of orthogonal groups, since
the smallest representation has index $2$ and hence $N$ is 
even.\footnote{This is true for 
$SO(n)$ groups with $n>6$.  Smaller $SO$ groups with $n=3,4,5,6$ are locally 
isomorphic to $SU(2)$, $SU(2)\times SU(2)$, $Sp(4)$, and $SU(4)$ 
groups, respectively, and the other constructions based on $SU$ or
$Sp$ groups apply.} 
This restores the continuous
$U(1)$ symmetry, which we can explicitly break to its $Z_N$ subgroup
by adding a mass term for the extra fermions $m\psi_0\psi_{-N}$ (or $m 
\psi_{-N/2}\psi_{-N/2}$ for $SO(n)$). If
$m$ is taken to be sufficiently big, it will influence neither the
low-energy theory nor the high-energy anomalies, 
but one can think of the mass parameter $m$ as a
spurion for breaking the continuous $U(1)$ symmetry to its $Z_N$
subgroup. Alternatively, for supersymmetric theories one can
promote the dynamical scale
$\Lambda^{b_0}=M^{b_0}e^{-(\frac{8\pi^2}{g^2(M)}+i\theta)}$ of the theory to a
background field with $U(1)$ charge $-\sum_i\mu_iq_i$, where 
$b_0$ is the coefficient of the one-loop $\beta$-function, $g$ is the
bare gauge coupling, $M$ is the ultraviolet cutoff, $\mu_i$ are the
Dynkin indices of the representations under the gauge group and $q_i$
are the $Z_N$ charges.  This restores 
the anomalous $U(1)$ symmetry because the effect of an
anomalous $U(1)$ symmetry is to shift the $\theta$ parameter, or in 
other words, a phase rotation of the scale $\Lambda$. One can
undo such a rotation by assigning the above charge under the $U(1)$
symmetry to the scale $\Lambda$~\cite{Seiberg}. 

By promoting the coupling constants of the theory to
background fields this way,
we have restored the continuous $U(1)$ global symmetries 
of the theory. In this theory the 't Hooft argument of Section~\ref{sec:tHooft}
holds, thus the anomaly matching conditions have to be satisfied for
this new $U(1)$ symmetry as well, together with all other continuous
global symmetries of the theory. Let us now consider what effect is 
generated to
the anomalies involving the broken $U(1)$ symmetry by
freezing the background fields to their actual value. For one, it breaks
the $U(1)$ to its $Z_N$ subgroup. However, since the background fields
do not carry $Z_N$ charge mod $N$, freezing of the background fields
does not change any of the anomalies mod $N$. Thus we conclude that
all the discrete anomalies involving the $Z_N$ discrete symmetry must
be matched mod $N$.  This argument however neglects four important
subtleties, which will change the final form of the discrete anomaly
matching conditions slightly: 
\begin{itemize}
\item decoupling of heavy fields
\item normalization of the $U(1)$ generators
\item charge fractionalization
\item different units of discrete charges for mixed $Z_N-Z_M$ anomalies.
\end{itemize}
In particular, charge fractionalization can invalidate the
discrete anomaly matching constraints for Type II anomalies, but not
for the Type I's.
In the following we describe the consequences of the above effects on
the discrete anomalies and then present the final form of the discrete
anomaly matching conditions explicitly. 

\subsubsection{Decoupling of Heavy Fermions}

As already mentioned at the end of Section~\ref{sec:inst}, the decoupling of
massive fermions can have non-trivial consequences on the discrete
anomaly matching conditions~\cite{IR}. 
The contribution of such particles to the  
continuous anomalies is always vanishing. 
This is not the case for discrete
anomalies and the possible contributions of such particles must be 
enumerated. One can, for example, have a pair of different Weyl fermions pairing 
up and acquiring a Dirac mass.
In this case, the charges of these fermions
must obey $q_1+q_2=mN$, where $q_1, q_2$, and $m$ are integers. These
fermions contribute integer multiples of $N$ to all anomalies, and since all
the anomaly matching equations are modulo $N$ anyway, such particles do
not change the anomaly matching 
equations. However for even $N$, there is another
possibility: a single
fermion with $Z_N$ charge $\frac{m}{2}N$ can acquire a Majorana mass. 
In this case,
the contribution of this fermion to the 
$Z_N({\rm gravity})^{2}$ anomaly is $\frac{m}{2}N$,
thus there could be a difference which is a half-integer multiple
of $N$ between the high-energy and the low-energy values of the
$Z_N({\rm gravity})^{2}$ anomaly. Similarly, the contribution
of such a Majorana fermion to the $Z_N^3$ anomalies is $N^3/8$. Thus
the $Z_N^3$ anomalies can differ by $mN^3/8$ if $N$ is even 
(as well as multiples of
$N$). The  $U(1)Z_N^2$ and $U(1)^2Z_N$ 
anomaly can not have a similar
contribution, since the $U(1)$ charge of a massive Majorana particle
must be zero. Similarly, it can not contribute to the $G_F^2Z_N$
anomaly either, since the Dynkin indices of real representations (as 
it is the case for Majorana fermions) are even. 

Therefore,
the possible existence of massive Majorana fermions leads to the
weaker anomaly matching condition for the $Z_N({\rm gravity})^{2}$ and the 
$Z_N^3$ anomalies. On the other hand, this also 
means that we might
gain some information (even if very limited) about the massive
spectrum as well. If the anomalies do differ by the additional factors
due to the Majorana fermions, we conclude 
that there must be odd number of massive Majorana fermions
with $Z_N$ charge $N/2$ present in the theory. In the case of anomaly
matching for dual pairs, an $N/2$ difference in the $Z_N({\rm
  gravity})^{2}$ anomaly signals that the number of
massive Majorana fermions with charge $N/2$ in the electric and
magnetic theories differs by an odd integer.

Furthermore, we can check the consistency of this assumption by
noting that there necessarily must be a difference of $N/2$ in the
$Z_N({\rm gravity})^{2}$ anomaly if there is a difference of $N^3/8$
in the $Z_N^3$ anomaly.\footnote{The converse is not
true. An $N/2$ difference in the $Z_N({\rm gravity})^{2}$ 
anomalies does not necessarily mean that an $N^3/8$
difference must be present in the $Z_N^3$ anomalies. The reason is
that if $N$ is divisible by four, then $N^3/8$ is automatically an
integer multiple of $N$, thus the effect of the decoupling Majorana
fermion can not be distinguished from the usual mod $N$ effects coming
from the non-uniqueness of the $Z_N$ charges. 
} Thus in addition to the fact
that the anomalies have to be matched, certain correlations among the
anomalies have to be satisfied as well.

\subsubsection{Normalization of the $U(1)$ Charges}

For the case of continuous anomaly matching conditions,
the overall normalization of the $U(1)$ charges  
is irrelevant. However for the discrete
$U(1)^2Z_N$ and $U(1)Z_N^2$ anomalies, this normalization is 
important since an overall change in the $U(1)$ charges can make
all equations mod $N$ to be satisfied. This is a valid argument for the
case of anomaly cancellation of gauged discrete symmetries~\cite{IR}. 
In the case of anomaly
matching, however, we do know the $U(1)$ charges of the high-energy
theory, and thus their normalization in the low-energy theory is fixed.
Then choosing a normalization in the high-energy theory such that all
$U(1)$ charges (including the ones in the low-energy theory) are integers
should result in valid $U(1)^2Z_N$ and $U(1)Z_N^2$ 
anomaly matching constraints. One needs to choose integer $U(1)$ charges;
otherwise a shift of $N$ in the $Z_N$ charges will not result in
shifts proportional to $N$ in the anomaly matching conditions. The
most stringent constraint arises, of course, if one chooses the
normalization of the $U(1)$ charges such that the charge assignments
are the smallest while they are still all integers.\footnote{If there
  are irrational $U(1)$ charges, we would not obtain any useful
  constraints.}

\subsubsection{Charge Fractionalization\label{fractional}}

In the context of anomaly cancellation for gauged discrete symmetries, 
Banks and Dine argued that the $Z_N^3$ anomalies do not lead to 
any condition on the low-energy theory~\cite{BD}. 
Their argument was that one can not decide whether one had really a
$Z_N$ or a $Z_{NM}$ symmetry in the high-energy theory from
the pure low-energy point of view.   
This could be the consequence of the fact that there are fractionally 
charged states in the high-energy theory but they decouple from the
low-energy theory. In particular, they argued that if there were 
states with charge $1/N$ in the high-energy theory, then at high energies
the $Z_N$ symmetry is enlarged to $Z_{N^2}$.  Since in the
low-energy theory all particles have charge zero mod $N$ under $Z_{N^2}$,
the $Z_{N^2}^3$ anomaly cancellation equations are trivially satisfied 
and hence give no useful information. In our case, however, the situation
is different. We know what the particle content of the high-energy theory
is, thus we know what the correct high-energy discrete symmetry group is.
Unless there are massive bound states or topological states in the theory
which carry fractional charges under the high-energy $Z_N$ symmetry, 
the $Z_N^3$ anomaly matching conditions must be satisfied as well. 
The situation is similar with the $U(1)Z_N^2$ and the $U(1)^2Z_N$
anomalies: if we assume
that the decoupled states carry integer $Z_N$ and $U(1)$ charges, then the 
$U(1)Z_N^2$ and $U(1)^2Z_N$ anomalies have to be matched mod $N$. In fact, 
in every example of anomaly matching that we considered 
and where all the other anomaly matching conditions were satisfied,
all the matching conditions for Type II anomalies ($Z_N^3$,
$U(1)^2Z_N$, $U(1)Z_N^2$, $U(1)_i U(1)_j Z_N$, $Z_N^2Z_M$ and
$U(1)Z_NZ_M$) were satisfied as well. This supports our claim that
the Type II 
anomaly matching conditions must be  considered as valid
constraints as well. However, we have to stress that the discrete
anomaly matching constraints for 
these Type II anomalies could in
principle be invalidated if charge fractionalization occurs for the
massive bound states, which can not be excluded on general grounds.
On the other hand, the Type I anomalies ($G_F^2Z_N$ and $Z_N({\rm
gravity})^2$) are not 
affected by a possible charge fractionalization and have to be always
matched.  

One can turn the above reasoning around for theories where one finds
that the Type I anomalies are matched while the Type II anomalies are
not matched, but there is ample of evidence for the considered
low-energy spectrum. In this case, the failure of the anomaly matching
for the Type II constraints could be used to gain some (even if very
limited) insight into the massive spectrum. We learn that there must be
massive states with fractional charges under the given symmetry for
which anomaly matching is not satisfied.

\subsubsection{Mixed $Z_N-Z_M$ Anomalies}

Finally, let us note that it is possible to have more than one discrete
symmetry in a theory, and that the discrete symmetry group is $Z_N\times
Z_M$. In this 
case, one has to consider the mixed $Z_N^2Z_M, Z_M^2Z_N$ and $Z_NZ_MU(1)_i$
anomalies as well.  Since the $Z_N$ charges are defined only
modulo $N$, while the $Z_M$ charges modulo $M$, the mixed anomalies can be
shifted by any integer combination of $N$ and $M$, $aN+bM$, where
$a,b$ are integers. If $N$ and $M$ are relatively prime, then $aN+bM$ can
take on any integer value and thus the mixed $Z_N-Z_M$ anomalies do not lead 
to any constraints. However if $N$ and $M$ have a common divisor $K$,
then $aN+bM$ is always a multiple of $K$. Thus in general the 
mixed $Z_N-Z_M$ anomaly matching conditions must hold modulo the greatest
common divisor of $N$ and $M$. If $N$ and $M$ are both even, then the
decoupling of massive Majorana fermions with charges $N/2,M/2$ can
yield an additional contribution of the form $N^2M/8$ to the
$Z_N^2Z_M$ anomalies, but there cannot be such contributions to 
the $Z_NZ_MU(1)$ anomalies.

\subsection{The Discrete Anomaly Matching Conditions\label{sec:recap}}

To summarize this section, we found that the presence of non-anomalous 
discrete global symmetries does yield anomaly matching constraints for these
theories. The anomalies have to match in the low-energy and high-energy
descriptions up to certain multiples of $N$. These anomalies and the possible
multiples of $N$ for the
different anomalies are given in Table~\ref{tab:summary}. 

\begin{table}
\begin{tabbing}
\hspace*{0.75cm} \= \hspace*{3cm}\=  
Anomaly \hspace*{2.5cm} \= Expression  \hspace*{2.25cm} 
\=   Difference  \\ \\ 
\> Type I \>  $G^2Z_N$: \hspace*{3cm} \> $\sum_i \mu_i q_i$ \hspace*{3cm} 
\>   $mN$ \\ 
\\
\> Type I \> $Z_N(\mbox{gravity})^{2}$: \> $\sum_i q_i$ \> $mN 
+\frac{m'}{2}N$ \\ \\
\> Type II \> $ Z_N^3$: \> $\sum_i q_i^3$ \> $mN + \frac{m'}{8}N^3$ \\  \\
\> Type II \> $ U(1)^2 Z_N$: \> $\sum_i Q_i^2q_i$ \> $mN$ \\  \\
\> Type II \> $ U(1)_i U(1)_j Z_N$: \> $\sum_i P_iQ_iq_i$ \> $mN$ \\  \\
\> Type II \> $ U(1)Z_N^2$: \> $\sum_i Q_iq_i^2$ \> $mN$ \\  \\
\> Type II \> $ Z_N^2Z_M$: \> $\sum_i q_i^2p_i$ \> $mK +\frac{m'}{8}N^2M$\\ \\
\> Type II \> $ U(1)Z_NZ_M$: \> $\sum_i Q_i q_i p_i$ \>  $mK$ \\
\end{tabbing}
\caption{The discrete anomaly matching conditions. The second column
  displays the given anomaly involving a discrete $Z_N$ symmetry. The
  third column gives the explicit expression how to evaluate this
  anomaly both in the high-energy and in the low-energy theories.  
  All charges are integers.  $\mu_{i}$ denotes the Dynkin index of 
  the representation $i$ under the non-Abelian group $G$.  The
  fourth column gives the allowed difference between the discrete anomalies
  evaluated in the high-energy and low-energy theories.  $m, m'$ 
  are integers, and $m'$ can be non-vanishing 
  only if $N,M$ are even.  $K$ is the GCD of $N$ and $M$. Type I
  anomaly matching constraints have to be satisfied regardless of
  the details of the massive spectrum. Type II anomalies have to be
  also matched except if there are fractionally charged
  massive states.
\label{tab:summary}}
\end{table}
In Table~\ref{tab:summary},
$m$ and $m'$ are integers, $K$ is the greatest common divisor (GCD)
of $N$ and
$M$, $q_i$ are $Z_N$ charges, $p_i$ are $Z_M$ charges, $Q_i$ and $P_i$ are
$U(1)$ charges, all the $q_i,p_i,Q_i,P_i$ are integers,
$G$ denotes a non-Abelian global symmetry, $\mu_i$ are the Dynkin
indices under this non-Abelian global symmetry.
$m'$ can be
non-zero only for $N,M$ even. The Type II 
anomaly matching conditions have to be matched as long as the massive 
spectrum carries integer charges.  They may be evaded if there are 
massive fractionally charged states.
The Type I constraints have to be always satisfied,
regardless of charge fractionalization.

\section{Discrete Anomalies and Seiberg Dualities\label{sec:SO}}
\setcounter{equation}{0}
\setcounter{footnote}{0}

As an application of the discrete anomaly matching conditions 
derived above, we show in this section that the exact  
results~\cite{Seiberg,IntrSeib,IntrPoul}
 on $N=1$ supersymmetric gauge theories indeed satisfy
these anomaly matching constraints. First of all, note that
SUSY $SU(N)$ QCD does not have a non-trivial discrete symmetry
besides the discrete subgroups of the continuous symmetries, as 
discussed in Section~\ref{sec:indep}.  The $Z_2$
color conjugation (see Section~\ref{sec:conjug}) 
exchanges the quarks and antiquarks, and is
hence not a useful symmetry for discrete anomaly matching either.
The same statement is true for the $Sp(2N)$ theories with
fundamentals, except that in these theories there is not even a color
conjugation present. Thus for the $SU$ and $Sp$ theories with only
fundamental representations, there is no discrete symmetry present to
check anomaly matching. The situation is different for $SO(N)$
theories with vectors. Since a vector of $SO(N)$ has Dynkin index two
({\it i.e.}\/ there are two zero modes for the vector in a one-instanton
background), there is a global $Z_{2F}$ symmetry which is not contained
in the continuous flavor symmetries as a discrete subgroup. Thus 
the global symmetries of this theory are:
\begin{equation}
\label{SOtable}
\begin{array}{c|cccc}
& SO(N)&SU(F)&U(1)_R & Z_{2F} \\ \hline
Q & \Yfund & \Yfund & 1-\frac{N-2}{F} & 1 \end{array}
\end{equation}

In addition to these symmetries, there is 
an extra $Z_2$ outer automorphism for $N=2n$ which is 
not part of the gauge or flavor symmetries, and the automorphism
defines a color conjugation symmetry as discussed in
Section~\ref{sec:discrete} and in \ref{app:charge}.  
The color conjugation can be defined to be 
the internal parity-like transformation
(color-parity ${\cal P}$),
which acts on vectors by 
flipping the sign of one particular color.\footnote{This is what 
Intriligator and Seiberg called ``charge conjugation'' ${\cal 
C}$~\cite{IntrSeib}.  
We, however, reserve the name ``charge conjugation'' only for 
the conventional ones given in Eq.~(\ref{cc}) to avoid confusions.} This
color-parity 
transformation for $N=4k+2$ is equivalent to the usual charge 
conjugation defined in Eq.~(\ref{cc}) up to a gauge transformation.  On 
the other hand, for $N=4k$ they are not equivalent since the charge 
conjugation is trivial up to a gauge transformation.  
Note that a similarly defined 
color-parity transformation for $SO(2n+1)$ theories is gauge equivalent to 
an overall $Z_2 \subset Z_{2F}$ global 
transformation and is hence of flavor-type; this is expected to be the 
case since there are no non-trivial outer automorphisms for 
$SO(2n+1)$.  However, for evaluating the discrete anomalies, we 
will use the color-parity transformation
${\cal P}$ for all $SO(N)$ groups so that the discrete anomalies are
given by the same expression regardless of $N$ being even or odd.

We have seen that the discrete symmetries of the $SO(N)$ theory
are $Z_{2F}\times {\cal P}$ for $N$ even and $Z_{2F}$ for odd $N$.
If $F$ is odd, then the $Z_{2F}$ symmetry is equivalent to a 
$Z_2\times Z_F$ symmetry.\footnote{In general, $Z_{NM}$ is equivalent
  to $Z_N\times Z_M$ if $N$ and $M$ are relatively prime.} 
However, the $Z_F$ factor is nothing but the center
of the $SU(F)$ flavor symmetry of the vectors, thus for odd $F$ the
non-trivial discrete symmetry of the theory is just the $Z_2$
sign flip of all vectors. If $N$ is even, this symmetry is
already contained in the gauge group, thus for odd $F$ even $N$, the
only discrete symmetry of the theory is ${\cal P}$. For odd $F$ 
odd $N$, the $Z_2$ sign flip of all vectors is not contained in the gauge
group, but there is no color conjugation (color-parity is gauge
equivalent to a flavor $Z_2$), thus the final symmetry of 
the theory is just $Z_2$. Therefore we conclude that the independent discrete
symmetries of the $SO(N)$ theory are:
\begin{eqnarray*}
&N \;\;\; {\rm even} \;\;\; F  \;\;\; {\rm even}: &Z_{2F}\times {\cal
  P}   \\
&N \;\;\; {\rm odd} \;\;\; F  \;\;\; {\rm even}: &Z_{2F}  \\
&N \;\;\; {\rm even} \;\;\; F  \;\;\; {\rm odd}: &{\cal P}  \\
&N \;\;\; {\rm odd} \;\;\; F  \;\;\; {\rm odd}: &Z_2
\end{eqnarray*}

We will write all anomaly matching conditions for the full
$Z_{2F}$ symmetry regardless of whether $F$ or $N$ is even or odd. Even if
$F$ is odd the $Z_{2F}$ anomaly matching conditions do have to be
satisfied; it is just that the $Z_F$ part of it  has to be
automatically 
satisfied due to the anomaly matching of the continuous $SU(F)$ 
symmetries, which Intriligator and Seiberg have already checked.

\subsection{$F>N-2$}
For $F>N-2$ the theory at the origin has 
a dual magnetic description in terms of the gauge group
$SO(F-N+4)$~\cite{IntrSeib}. The global symmetries of the dual theory are:
\begin{equation} \begin{array}{c|cccc}
& SO(F-N+4) & SU(F) & U(1)_R & Z_{2F} \\ \hline
q & \Yfund & \overline{\Yfund} & \frac{N-2}{F} & -1 \\
M & 1 & \Ysymm & 2-2\frac{N-2}{F} & 2 \end{array}, \end{equation}
and there is also a superpotential $W_{mag}=Mq^2$ in the magnetic
theory.\footnote{For $F=N-1$, there is an additional $W={\rm det}\, M$
  term in the magnetic superpotential. The presence of this extra 
  term, however, does not affect anomaly matching.}
The $Z_{2F}$ charge of $M$ is determined by the matching
$Q^2\leftrightarrow M$, while the $Z_{2F}$ charge of $q$ is determined
by the requirement that the superpotential has zero $Z_{2F}$ charge 
mod $2F$. Note that this does not completely fix the charge of $q$, since
one could as well add $F$ to it. This modification does indeed happen, but
in a very subtle way. 
It has been already noted in~\cite{IntrSeib} that the mapping
of baryon operators implies a non-trivial mapping between the
$Z_{2F}$ symmetries of the electric and the magnetic theories. 
The baryon $Q^N$ of the electric theory is mapped to the
``exotic baryon'' $\tilde{W}_{\alpha}\tilde{W}^{\alpha}q^{F-N}$ of the magnetic 
theory.  (Both of them have all color indices contracted with the 
$\epsilon$-tensor.)
Comparing the phases of these operators under a $Z_{2F}$ symmetry
transformation, we see that there is an overall sign difference in the
$Z_{2F}$ transformation properties of the two operators. The
resolution of this puzzle is mixing of the $Z_{2F}$ symmetry with
the color-parity transformation ${\cal P}$. The
$Z_{2F}$ symmetry of the electric theory is mapped to ${\cal P}Z_{2F}$ of the
magnetic theory, $Z_{2F}\leftrightarrow {\cal P} Z_{2F}$. 
This takes care of the difference in the sign of the baryon under 
$Z_{2F}$ transformation, since the baryon operators have all color 
indices contracted by $\epsilon$-tensors, and the effect of the color-parity
transformation is to flip the sign of one particular color.

To see the mapping of discrete symmetries more precisely, 
we have to separate the cases when $N$ and $F$ are
even or odd. If both $N$ and $F$ are even, the electric theory
has a $Z_{2F}\times {\cal P}$ discrete symmetry.  Since in this case 
$\tilde{N}$ and $F$ are both even as well ($\tilde{N}=F-N+4$, the size
of the dual gauge group), the magnetic theory also has a  $Z_{2F}\times {\cal
P}$ discrete symmetry, and the mapping of the discrete 
symmetries is as above:  $Z_{2F}\leftrightarrow {\cal P} Z_{2F}$, 
while $ {\cal P}\leftrightarrow  {\cal P}$. If $N$ is even and $F$ is odd the
electric theory has only the ${\cal P}$ symmetry,
while $\tilde{N}$ and $F$ in the magnetic theory are both odd, and
thus the magnetic theory has only a $Z_2$ discrete symmetry, and the
mapping is given by  $ {\cal P}\leftrightarrow Z_2$. If $N$ is odd and $F$ is
even, the electric theory has a $Z_{2F}$ symmetry.  Since 
$\tilde{N}$ is odd and $F$ is even in the magnetic theory, the
magnetic theory also has a $Z_{2F}$ symmetry, with the mapping
$Z_{2F}\leftrightarrow Z_{2F}$ (the generator $\omega = e^{2\pi i/2F}$
of $Z_{2F}$ is, however,
mapped to $-\omega$). Finally, for both $N$ and $F$ odd,
the electric theory has a $Z_2$ symmetry.  Since $\tilde{N}$ is
even and $F$ is odd, the magnetic theory has the
${\cal P}$ color-parity symmetry, and the mapping of symmetries
is given by $Z_2\leftrightarrow {\cal P}$. Therefore, we find that the
discrete global 
symmetries of the electric and the magnetic theories match for every
possible combination of parities of $N$ and $F$, and the
mapping of the discrete symmetries is given above. 

In the following, we show that the anomalies involving 
the $Z_{2F}$ discrete symmetries match using the above mapping of the discrete
symmetries. Regardless of whether $F$ and $N$ are even or odd, we will
calculate the anomalies for the full $Z_{2F}$ group, and use the same
mapping of symmetries. For some particular cases this symmetry may involve a
piece for which anomaly matching follows from the continuous anomalies,
but we never lose any information by considering the bigger group.

The effect of the color-parity transformation
in the mapping of the $Z_{2F}$ symmetry has
to be taken into account when one compares the 
discrete anomalies of the electric and the magnetic theories. This can be
done by adding an extra $Z_{2F}$ charge $F$ to every field that
carries the first magnetic color: $q^1$ has $Z_{2F}$
charge $F-1$, and the gluinos $\tilde{\lambda}^{1i}$ $(i\neq 1)$ have $Z_{2F}$
charge $F$.  With this knowledge at hand, we can 
check the discrete anomaly matching conditions.  In the following
list of anomalies, we write contributions to the anomalies in the
magnetic theory in the order of $q^1$, $q^i$ $(i\neq 1)$, $M$ and
$\tilde{\lambda}^{1i}$.

\begin{tabbing}
\hspace*{1cm} \= \hspace*{2.5cm} \= 
\hspace*{1.5cm} Electric theory \hspace*{0.5cm} 
\= \hspace*{1cm} Magnetic theory \\
\>  \begin{minipage}[c]{2cm} \[ SU(F)^2Z_{2F}\] \end{minipage} 
\> \begin{minipage}[c]{6cm} \[ N \] \end{minipage} 
\> \begin{minipage}[c]{5cm}
\begin{eqnarray*} && (F-1)-(F-N+3)+\\ &&2(F+2)=N+2F
\end{eqnarray*} \end{minipage} \\

\>  \begin{minipage}[c]{2cm} \[ Z_{2F}(\mbox{gravity})^{2} \] \end{minipage} 
\> \begin{minipage}[c]{6cm} \[ NF \] \end{minipage} 
\> \begin{minipage}[c]{6cm}
\begin{eqnarray*} && F(F-1)-F(F-N+3)+  \\
&& F(F+1)+(F-N+3)F= \\ &&2F^2 \end{eqnarray*} 
\end{minipage} \\
\>  \begin{minipage}[c]{2cm} \[ Z_{2F}^3\] \end{minipage} 
\> \begin{minipage}[c]{6cm} \[ NF \] \end{minipage} 
\> \begin{minipage}[c]{6cm}
\begin{eqnarray*} && F(F-1)^3-F(F-N+3)+  \\ && 4F(F+1)+
F^3(F-N+3)=  \\ &&  NF-
NF^3 \; {\rm mod} \; 2F  \end{eqnarray*} 
\end{minipage} \\

\>  \begin{minipage}[c]{2cm} \[ U(1)_R^2 Z_{2F}\] \end{minipage} 
\> \begin{minipage}[c]{6cm} \begin{eqnarray*} && NF(N-2)^2=\\ 
 &&N^3F  \; {\rm mod} \; 2F
 \end{eqnarray*} \end{minipage} 
 \> \begin{minipage}[c]{6cm}
\begin{eqnarray*} && F(N-2-F)^2(F-1)- \\
&& (F-N+3)F(N-2-F)^2+ \\
&& F(F+1)(F-2N+4)^2+  \\
&& F^3(F-N+3) =  \\
&&   N^3F  \; {\rm mod} \; 2F  \end{eqnarray*} 
\end{minipage} \\

\>  \begin{minipage}[c]{2cm} \[ U(1)_R Z_{2F}^2\] \end{minipage}
\> \begin{minipage}[c]{6cm} \begin{eqnarray*} &&-NF(N-2)=\\
 &&-N^2F  \; {\rm mod} \; 2F
 \end{eqnarray*} \end{minipage} 
\> \begin{minipage}[c]{6cm}
\begin{eqnarray*} && F(N-2-F)(F-1)^2+  \\
&& F(F-N+3)(N-2-F)+ \\
&& 2F(F+1)(F-2N+4)+ \\
&& (F-N+3)F^3= \\
&& -N^2 F  \; {\rm mod} \; 2F  \end{eqnarray*} 
\end{minipage} \\
\end{tabbing}

The $SU(F)^2Z_{2F}$, $U(1)_R^2Z_{2F}$ and $U(1)_RZ_{2F}^2$ anomalies
obviously match 
${\rm mod}\;  2F$. The $Z_{2F}({\rm gravity})^2$ anomalies match only 
${\rm mod}\;  F$ for odd $N$, which signals that the difference of the
number of massive Majorana fermions with $Z_{2F}$ charge $F$ in the electric
and magnetic theories is odd.
This is confirmed by the fact that there is a term
$-NF^3$ appearing in the $Z_{2F}^3$ anomalies, which is of the
form $m(\frac{2F}{2})^3$, and can again be attributed to the
decoupling 
of Majorana particles with $Z_{2F}$ charge $F$. Thus these anomalies
obey the discrete anomaly matching conditions as well.  Therefore, all the
discrete anomaly matching conditions are satisfied in a rather
non-trivial way. Note that we chose
$U(1)_R$ charges that are $F$-times the charges in Table~\ref{SOtable}
for the anomalies involving $U(1)_R$, 
in order to obtain integer $U(1)_R$ charge assignments for all
fields.

\subsection{$F=N-2$}
In the case of $F=N-2$, the theory is in the Abelian Coulomb phase, with $F$ 
pairs of magnetic monopoles becoming massless at the
origin~\cite{IntrSeib}. 
Thus the field content of the low-energy theory is given by
\begin{equation} \begin{array}{c|cccc}
& U(1) & SU(N-2) & U(1)_R & Z_{2N-4} \\ \hline
M & 0 & \Ysymm & 0 & 2 \\
q^+ & 1 & \overline{\Yfund} & 1 & -1 \\
q^- & -1 & \overline{\Yfund} & 1 & -1 \end{array}, \end{equation}
and a superpotential $W=Mq^+q^-$.
There are no baryons in either the high-energy or the low-energy
theory, while the ``exotic baryon'' $W_{\alpha} Q^{N-2}$ of the
original $SO(N)$ theory is mapped to the photon $\tilde{W}_{\alpha}$ 
of the low-energy $U(1)$ theory~\cite{IntrSeib}. Since there is again a 
sign difference in the $Z_{2N-4}$ transformation properties of these two 
operators, the $Z_{2N-4}$ symmetry of the $SO(N)$ theory is 
mapped to ${\cal P}Z_{2N-4}$ in the $U(1)$ theory. This can be taken into 
account by adding $N-2$ to the $Z_{2N-4}$ charge of one of the 
monopoles (say $q^{1} = i(q^+-q^-)/\sqrt{2}$ which corresponds 
to the first $SO(2)$ color)
and of the photino $\tilde{\lambda}$. Now we can calculate the discrete
anomalies in both the high-energy $SO(N)$ theory and the 
low-energy $U(1)$ description.  In the following
list of anomalies, we write contributions to the anomalies in the
magnetic theory in the order of $q^1$, $q^2$, $M$ and
$\tilde{\lambda}$.  

\begin{tabbing}
\= \hspace*{3cm} \= \hspace*{2.5cm} UV \hspace*{2cm} 
\= \hspace*{3cm}IR \hspace*{1cm} \\
\>  \begin{minipage}[c]{2cm} \[ SU(N-2)^2Z_{2N-4}\] \end{minipage} 
\> \begin{minipage}[c]{6cm} \[ N \] \end{minipage} 
\> \begin{minipage}[c]{5cm}
\begin{eqnarray*} && (N-3)-1+2N=3N-4\end{eqnarray*} \end{minipage} \\

\>  \begin{minipage}[c]{2cm} \[ Z_{2N-4}(\mbox{gravity})^{2} \] \end{minipage} 
\> \begin{minipage}[c]{6cm} \[ N(N-2) \] \end{minipage} 
\> \begin{minipage}[c]{6cm}
\begin{eqnarray*} && (N-2)(N-3)-(N-2)+  \\
&& (N-2)(N-1)+(N-2)=  \\
&& (N-2)(2N-4)  
\end{eqnarray*} \end{minipage} \\

\>  \begin{minipage}[c]{2cm} \[ Z_{2N-4}^3\] \end{minipage} 
\> \begin{minipage}[c]{6cm} \[ N(N-2) \] \end{minipage} 
\> \begin{minipage}[c]{6cm}
\begin{eqnarray*} && (N-2)(N-3)^3-(N-2) + \\
&& 4(N-2)(N-1)+(N-2)^3 =  \\
&& N(N-2)^3+N(N-2) \; {\rm mod} \; 2N-4  \end{eqnarray*} 
\end{minipage} \\

\>  \begin{minipage}[c]{2cm} \[ U(1)_R^2 Z_{2N-4}\] \end{minipage} 
\> \begin{minipage}[c]{6cm} \[ N(N-2)^3 
 \] \end{minipage} 
\> \begin{minipage}[c]{6cm}
\begin{eqnarray*} && (N-2)(N-1)+(N-2)=  \\
&&   N(N-2)  \end{eqnarray*} 
\end{minipage} \\

\>  \begin{minipage}[c]{2cm} \[ U(1)_R Z_{2N-4}^2\] \end{minipage}
\> \begin{minipage}[c]{6cm} \[ -N(N-2)^2 \] \end{minipage} 
\> \begin{minipage}[c]{6cm}
\begin{eqnarray*} && -2(N-2)(N-1)+(N-2)^2= \\
&& (N-2)^2 \; {\rm mod} \; 2N-4  \end{eqnarray*} 
\end{minipage} \\
\end{tabbing}

The $ SU(N-2)^2Z_{2N-4}$ anomalies obviously match mod $2N-4$. 
The $Z_{2N-4}(\mbox{gravity})^{2}$ anomalies match only modulo 
$N-2$ for odd $N$, which signals the presence of massive Majorana fermions
with charge $N-2$. The difference in the $Z_{2N-4}^3$ anomalies
is $N(N-2)^3$ modulo $2N-4$. The $(N-2)^3$ term is 
due to the presence of the massive Majorana fermions.
The difference in the $ U(1)_R^2 Z_{2N-4}$ anomalies is 
$(N-2)N[(N-2)^2-1]$, which is a multiple of $2N-4$ since $N[(N-2)^2-1]$ is
even. Similarly, the difference $(N-2)^2(1+N)$
in the $ U(1)_R Z_{2N-4}^2$ anomalies is a multiple of $2N-4$ since
$(N-2)(1+N)$ is even. Thus all discrete anomaly matching conditions 
for the $F=N-2$ case are satisfied.

\subsection{$F=N-3$}
For $F=N-3$, there are two branches of the theory~\cite{IntrSeib}. 
On one branch, there
is a dynamically generated superpotential and we do not expect
anomaly matching. On the other branch, however, the
theory close to the origin is described by massless gauge singlet
composites $M$ and $b$, whose global symmetry properties are
\begin{equation} \begin{array}{c|ccc}
& SU(N-3) & U(1)_R & Z_{2N-6} \\ \hline
M & \Ysymm & \frac{-2}{N-3} & 2 \\
b & \overline{\Yfund} & 1+\frac{1}{N-3} & N-4 \end{array}
\end{equation}
and there is a superpotential $W=Mb^2$ in the low-energy theory.
The field
$b$ can be identified with the exotic baryon $W_{\alpha}W^{\alpha}Q^{N-4}$
of the original $SO(N)$ theory. The $Z_{2N-6}$ charges of $M$ and $b$
have  been
chosen such that this mapping is exactly obeyed. In the following
list of anomalies, we write contributions to the anomalies in the
magnetic theory in the order of $M$ and $b$.  
The discrete anomalies are:

\begin{tabbing}
\= \hspace*{3cm} \= \hspace*{2.5cm} UV \hspace*{2cm} 
\= \hspace*{3cm}IR \hspace*{1cm} \\
\>  \begin{minipage}[c]{2cm} \[ SU(N-3)^2Z_{2N-6}\] \end{minipage} 
\> \begin{minipage}[c]{6cm} \[ N \] \end{minipage} 
\> \begin{minipage}[c]{6cm}
\begin{eqnarray*} &&2(N-1)+(N-4)=3N-6\end{eqnarray*} \end{minipage} \\

\>  \begin{minipage}[c]{2cm} \[ Z_{2N-6}(\mbox{gravity})^{2} \] \end{minipage} 
\> \begin{minipage}[c]{6cm} \[ N(N-3) \] \end{minipage} 
\> \begin{minipage}[c]{6cm}
\begin{eqnarray*} && (N-2)(N-3)+(N-4)(N-3)=  \\
&& (N-3)(2N-6)  
\end{eqnarray*} \end{minipage} \\

\>  \begin{minipage}[c]{2cm} \[ Z_{2N-6}^3\] \end{minipage} 
\> \begin{minipage}[c]{6cm} \[ N(N-3) \] \end{minipage} 
\> \begin{minipage}[c]{6cm}
\begin{eqnarray*} && 4(N-2)(N-3)+ \\
&& (N-3)(N-4)^3=  \\
&& (N-3)(N-4)^3 \; {\rm mod} \; 2N-6  \end{eqnarray*} 
\end{minipage} \\

\>  \begin{minipage}[c]{2cm} \[ U(1)_R^2 Z_{2N-6}\] \end{minipage} 
\> \begin{minipage}[c]{6cm} \[ N(N-2)^2(N-3) 
 \] \end{minipage} 
\> \begin{minipage}[c]{6cm}
\begin{eqnarray*} && (1-N)^2(N-3)(N-2)+  \\
&& (N-3)(N-4)=  \\
&&   (N-3)(N-4) \; {\rm mod} \; 2N-6  \end{eqnarray*} 
\end{minipage} \\

\>  \begin{minipage}[c]{2cm} \[ U(1)_R Z_{2N-6}^2\] \end{minipage}
\> \begin{minipage}[c]{6cm} \[ -N(N-2)(N-3) \] \end{minipage} 
\> \begin{minipage}[c]{6cm}
\begin{eqnarray*} && 2(N-2)(N-3)(1-N)+  \\
&& (N-3)(N-4)^2 = \\
&& (N-3)(N-4)^2 \; {\rm mod} \; 2N-6  \end{eqnarray*} 
\end{minipage} \\
\end{tabbing}
The $SU(N-3)^2Z_{2N-6}$ anomaly is obviously matched modulo $2N-6$. The
$Z_{2N-6}(\mbox{gravity})^{2}$ anomaly is matched modulo $2N-6$ if $N$ is even
and modulo $N-3$ if $N$ is odd. Thus in the odd $N$ case we again have
massive Majorana particles carrying half of the total $Z_{2N-6}$
charge. However such particles will not affect the $Z_{2N-6}^3$ 
anomalies, since for odd $N$ the term $(2N-6)^3/8$ is a multiple
of $2N-6$. As expected from this argument, the $Z_{2N-6}^3$ 
anomalies match modulo $2N-6$, without extra cubic contributions.
Finally, the difference $(N-3)[(N-4)-N(N-2)^2]$ in the
 $U(1)_R^2 Z_{2N-6}$ anomalies and $(N-3)[(N-4)^2+N(N-2)]$ for the
 $U(1)_R Z_{2N-6}^2$ anomalies are both multiples of $2N-6$. 
Thus we have again found that all discrete anomalies are matched 
in the $F=N-3$ case.

\subsection{$F=N-4$\label{sec:N-4}}

For $F=N-4$, there are again two branches~\cite{IntrSeib}. 
On one branch there is a 
dynamically generated superpotential and this branch is not of
our interest. On the other branch there is a moduli space of 
quantum vacua described by the meson field $M$ and no superpotential.
The global symmetry properties of $M$ are given by
\begin{equation} \begin{array}{c|ccc}
& SU(N-4) & U(1)_R & Z_{2N-8} \\ \hline
M & \Ysymm & -\frac{4}{N-4} & 2 \end{array}. \end{equation}
While all the continuous anomalies are matched by $M$, the anomalies
involving the discrete $Z_{2N-8}$ symmetries are not matched. For example,
the $SU(N-4)^2Z_{2N-8}$ anomaly is $N$ in the original $SO(N)$ theory
and $2N-4$ in the infrared; the difference is not a multiple
of $2N-8$. 

The resolution to this puzzle is that the discrete
$Z_{2N-8}$ symmetry is actually spontaneously broken to $Z_{N-4}$. 
To understand
this breaking of the $Z_{2N-8}$ symmetry, we have to consider the 
details of the dynamics of the theory. On a generic point in the
moduli space, where all $N-4$ $Q$'s have expectation values, the
$SO(N)$ gauge group is broken to $SO(4)\sim SU(2)_L\times SU(2)_R$.
The matching of the scales is given by 
$\Lambda_L^6=\Lambda_R^6=\Lambda^{2(N-1)}/({\rm det} M)$.
Then gaugino condensation occurs in both $SU(2)$ gauge groups,
producing the superpotential
\begin{equation} \frac{1}{2} (\epsilon_L+\epsilon_R)\left( \frac{16\Lambda^{2(N-1)}}{{\rm
det 
}M}\right)^{\frac{1}{2}}, \end{equation}
where $\epsilon_{L,R}$ are $\pm 1$. The branch with $\epsilon_L\epsilon_R=1$
corresponds to the theory with a dynamical superpotential. The 
branch with $\epsilon_L\epsilon_R=-1$ produces the theory with no 
superpotential and with a moduli space of vacua. Even 
though there is no superpotential generated, the existence of a gaugino 
condensate 
already suggests that some of the global symmetries might be
spontaneously broken. A pure super Yang-Mills theory has only
discrete symmetries, so we expect that the broken symmetry
is only the $Z_{2N-8}$ discrete symmetry. 

To see the spontaneous breakdown of the discrete symmetry
explicitly, we have to identify the symmetry properties of the
$SU(2)$ gauginos. The glueball field $S=W_{\alpha}^{SU(2)}W^{\alpha ,SU(2)}$
can be identified with the exotic composite baryon
$S\leftrightarrow W_{\alpha}W^{\alpha} Q^{N-4}$ of the $SO(N)$ 
theory. The transformation properties of this operator under the
global symmetries are:

\begin{equation} \begin{array}{c|ccc}
& SU(N-4) & U(1)_R & Z_{2N-8} \\ \hline
S & 1 & 0 & N-4 \end{array} \qquad .
\end{equation}
Thus one can see that the effect of the expectation value to $S$
is to leave all continuous global symmetries unbroken, but to break the
discrete $Z_{2N-8}$ symmetry to its $Z_{N-4}$ subgroup. Therefore in this 
case,
one only has to check the anomaly matching conditions with respect to this
$Z_{N-4}$ discrete group. These anomaly matching conditions, however, are 
all automatically satisfied, since
this $Z_{N-4}$ group can be identified with the 
center of the global $SU(N-4)$ symmetry whose anomaly matching is 
already checked. An explicit calculation of
the anomalies confirms this result. 

A method to verify the expectation value of the $S$ field 
is to first add another flavor and decouple it with a mass term.  On 
the branch in the $F=N-3$ theory without a runaway
superpotential, the 
superpotential is given by $W = M^{ij} b_{i} b_{j} - m M^{N-3,N-3}$.  
The equation of motion 
for $M^{N-3,N-3}$ gives $b_{N-3} = 
\sqrt{m}$.  Recall that $b_{N-3} = W_{\alpha} W^{\alpha} Q^{N-4}$ and 
is nothing but the field $S$ above.

To summarize this section, we have shown that the Seiberg results on
$N=1$ supersymmetric gauge theories all satisfy the discrete anomaly
matching conditions of the previous section. In the case of $SU$ and
$Sp$ theory this is not new, since all the discrete symmetries are 
subgroups of the continuous symmetries, and the anomaly matching conditions
follow from those of the continuous symmetries. However for $SO(N)$ 
groups,
the discrete symmetries are not all contained in the continuous global
symmetries.  We have seen that the anomaly matching conditions are
rather non-trivial, and give us further confidence in both Seiberg's 
results as well as in the method of discrete anomaly matching described in the
previous section.

\section{More $N=1$ Supersymmetric Examples\label{sec:other}}
\setcounter{equation}{0}
\setcounter{footnote}{0}

In this section we present several other examples of discrete anomaly
matching conditions for $N=1$ supersymmetric theories. First we
present two s-confining $SO$ theories~\cite{sconf}. 
In both of these examples, the origin of the discrete symmetry is the
higher Dynkin index of the representation. 

Then we present an $SU(6)$
example with a three-index antisymmetric tensor~\cite{sconf}. If there are
three additional flavors of $SU(6)$ fundamentals 
present in the theory, the $SU(6)$ theory
confines with a quantum deformed moduli space which breaks the global 
continuous symmetries spontaneously.  On one point of the
moduli space, the global continuous 
symmetries leave an unbroken $Z_{12}$ discrete
symmetry, and we show that the matching of this symmetry is
satisfied. If there are no flavors present, the theory is claimed to
have multiple branches \cite{sconf}, with one branch having a dynamical
superpotential while the other branch with a moduli space of vacua.  
The matching of discrete anomalies does not appear to work on the 
moduli space.  We
show how this second branch arises and also show that the discrete 
$Z_6$ global symmetry is actually spontaneously broken to $Z_2$, and 
hence the puzzle is resolved.

Next we consider the ``ISS-model'': $SU(2)$ with a three-index
symmetric tensor. This theory was argued to confine and break
supersymmetry after an appropriate tree-level superpotential is
added~\cite{ISS}. We show that in this theory the discrete anomaly
conditions are satisfied as well, giving additional (weak)
evidence in favor of the
description of Ref.~\cite{ISS}.

Finally, we
present the most non-trivial example of discrete anomaly matching that
we found. It is based on the Kutasov-type duality of Ref.~\cite{Intrdual} 
for $SO(N)$ with a symmetric tensor and additional vectors under
$SO(N)$. This theory has two different discrete symmetries: one from
the explicit breaking of a global $U(1)$ by the tree-level
superpotential term, while the other from the breaking of the
anomalous $U(1)$ due to instantons. We show how these symmetries are
mapped to the dual theory and that all anomaly matching
conditions are satisfied. 

\subsection{S-confining Theories}
\subsubsection{$SO(9)$ with Four Spinors}
The first s-confining example we present is $SO(9)$ with four
spinors~\cite{sconf}. Since the Dynkin index of the spinor is four, there is a
discrete global $Z_{16}$ symmetry in this theory. The global
symmetries of the theory together with the conjectured confined
low-energy bound states is given in the table below.

\begin{equation} \begin{array}{c|c|ccc}
& SO(9) & SU(4) & U(1)_R & Z_{16}\\ \hline 
S & 16 & \Yfund & \frac{1}{8} & 1 \\ \hline \hline
S^2 & & \Ysymm & \frac{1}{4} & 2 \\
S^4 & & \Ysquare & \frac{1}{2} & 4 \\
S^6 & & \Ysymm & \frac{3}{4} & 6 \end{array} \end{equation}

The anomaly matching conditions are:
\begin{tabbing}
\hspace*{1.5cm} \= \hspace*{3cm} \= \hspace*{1.5cm} UV \hspace*{1cm} 
\= \hspace*{3cm}IR \hspace*{1cm} \\
\> \begin{minipage}[c]{2cm} \[  SU(4)^2Z_{16} \] \end{minipage}
\> \begin{minipage}[c]{4cm} \[ 16  \] \end{minipage} 
\> \begin{minipage}[c]{8cm}
\[ 2\times 6 + 4\times 16 + 6\times 6 =112 \] \end{minipage} \\
\>  \begin{minipage}[c]{2cm} \[ Z_{16}(\mbox{gravity})^{2} \] 
\end{minipage} 
\> \begin{minipage}[c]{4cm} \[ 64 \] \end{minipage} 
\> \begin{minipage}[c]{8cm}
\[ 2\times 10+4\times 20+6\times 10=160 \]
\end{minipage} \\
\>  \begin{minipage}[c]{2cm} \[ Z_{16}^3\] \end{minipage} 
\> \begin{minipage}[c]{4cm} \[ 64 \] \end{minipage} 
\> \begin{minipage}[c]{8cm}
\[ 2^{3}\times 10+4^{3}\times 20+6^{3}\times 10=220\times 16\]
\end{minipage} \\
\>  \begin{minipage}[c]{2cm} \[ U(1)_R^2 Z_{16}\] 
\end{minipage} 
\> \begin{minipage}[c]{4cm} \[ 196 \times 16
 \] \end{minipage} 
\> \begin{minipage}[c]{8cm}
\[ 140 \times 16 \] 
\end{minipage} \\
\vspace*{-2cm} \>  \begin{minipage}[c]{2cm} \[  U(1)_R Z_{16}^2\] 
\end{minipage}
\> \begin{minipage}[c]{4cm} \renewcommand{\baselinestretch}{2.0}
\[ -28\times 16 \] \end{minipage} 
\> \begin{minipage}[c]{8cm} \[ -196\times 16 \]
\end{minipage} \\
\end{tabbing}
where the contributions to the first three anomalies in the magnetic theory 
are quoted in the order $S^{2}$, $S^{4}$, $S^{6}$.  The $U(1)_{R}$ 
charges are multiplied by a factor of 8 to make all the charges 
integers.
One can see that all anomalies match mod $16$.

\subsubsection{$SO(7)$ with Six Spinors}
The global symmetries of the theory and the low-energy confining
spectrum is given in the table below~\cite{sconf}.

\begin{equation} \begin{array}{c|c|ccc}
& SO(7) & SU(6) & U(1)_R & Z_{12} \\ \hline
S & 8 & \Yfund & \frac{1}{6} & 1 \\ \hline \hline
S^2 & & \Ysymm & \frac{1}{3} & 2 \\
S^4 & & \overline{\Yasymm} & \frac{2}{3} & 4 \end{array} \end{equation}

The anomaly matching conditions are:

\begin{tabbing}
\hspace*{1.5cm}\= \hspace*{3cm} \= \hspace*{2.5cm} UV \hspace*{2cm} 
\= \hspace*{3cm}IR \hspace*{1cm} \\
\>  \begin{minipage}[c]{2cm} \[ SU(6)^2Z_{12} \] \end{minipage} 
\> \begin{minipage}[c]{6cm} \[ 8  \] \end{minipage} 
\> \begin{minipage}[c]{6cm}
\[ 2\times 8+4\times 4=8+2\times 12 \] \end{minipage} \\
\>  \begin{minipage}[c]{2cm} \[ Z_{12}(\mbox{gravity})^{2} \] \end{minipage} 
\> \begin{minipage}[c]{6cm} \[ 48 \] \end{minipage} 
\> \begin{minipage}[c]{6cm}
\[  2\times 12+4\times 15 = 8\times 12 + 6\]
\end{minipage} \\
\>  \begin{minipage}[c]{2cm} \[ Z_{12}^3\] \end{minipage} 
\> \begin{minipage}[c]{6cm} \[ 48 \] \end{minipage} 
\> \begin{minipage}[c]{6cm}
\[ 2^{3}\times 21+4^{3}\times 15=94\times 12\]
\end{minipage} \\
\>  \begin{minipage}[c]{2cm} \[ U(1)_R^2 Z_{12}\] \end{minipage} 
\> \begin{minipage}[c]{6cm} \[ 1200
 \] \end{minipage} 
\> \begin{minipage}[c]{6cm}
\[ 76\times 12\] 
\end{minipage} \\
\>  \begin{minipage}[c]{2cm} \[  U(1)_R Z_{12}^2\] \end{minipage}
\> \begin{minipage}[c]{6cm} \[ -5\times 8\times 6 \] \end{minipage} 
\> \begin{minipage}[c]{6cm} \[ -68\times 12 \]
\end{minipage} \\
\end{tabbing}
where the contributions to the first three anomalies in the magnetic theory 
are quoted in the order $S^{2}$, $S^{4}$.  The $U(1)_{R}$ 
charges are multiplied by a factor of 6 to make all the charges 
integers.

All anomalies match mod $12$ except the $Z_{12}(\mbox{gravity})^{2}$
anomaly, which is matched mod $6$, and signals the presence of
massive Majorana fermions with charge $6$.  But we do not see the 
corresponding contribution to $Z_{12}^{3}$ anomaly because $12^{3}/8 = 
216$ is a multiple of 12.  

\subsection{Quantum Modified Constraint and Moduli Space of Vacua}
Next we present two examples using an $SU(6)$ theory with a 
three-index antisymmetric tensor and fundamental
flavors~\cite{sconf}. The first
example is $SU(6)$ with $\Ythreea$ and $3(\Yfund
+\overline{\Yfund})$. This theory is confining with one quantum modified
and one unmodified constraint. The matter fields, global symmetries
and the confining spectrum of the theory are:
\begin{displaymath}
\label{su6}
\begin{array}{c|c|ccccc} 
 & SU(6) & SU(3)_{Q}& SU(3)_{\bar{Q}}& U(1)_B & U(1)_A & U(1)_R \\ \hline  
 A & \Ythreea & 1 & 1 & 0 & 1 & 0 \\
 Q & \Yfund & \Yfund & 1 & 1 & -1 & 0 \\
 \bar{Q} & \overline{\Yfund} & 1 & \Yfund & -1 & -1 & 0 \\ \hline \hline
M_0=Q\bar{Q}& & \Yfund & \Yfund & 0 & -2 & 0 \\
M_2=QA^2\bar{Q} & & \Yfund & \Yfund & 0 & 0 & 0 \\
B_1=AQ^3 & & 1 & 1 & 3 & -2 & 0 \\
\bar{B}_1 =A \bar{Q}^3 & & 1 & 1 & -3 & -2 & 0 \\
B_3=A^3 Q^3 & & 1 & 1 & 3 & 0 & 0 \\
\bar{B}_3=A^3 \bar{Q}^3 & & 1 & 1 & -3 & 0 & 0 \\
T=A^4 & & 1 & 1 & 0 & 4 & 0 \end{array}
\end{displaymath}

The superpotential implementing the constraints is
\begin{eqnarray}
W&=& \lambda \left(B_1 \bar{B}_1 T + B_3 \bar{B}_3 + M_2^3 + 
             T M_2 M_0^2 -
             \Lambda^{12} \right) +  \nonumber \\
  &&         \mu \left(M_2^2 M_0 + T M_0^3 +
             \bar{B}_1 B_3 + B_1 \bar{B}_3 \right),
\end{eqnarray}
where $\lambda$ and $\mu$ are Lagrange multipliers. The original high-energy 
theory does not have an interesting discrete symmetry. One
might think that there is a $Z_6$ symmetry rotating only the $A$
field. However, if one in addition to this $Z_6$ performs a
discrete $U(1)_A$ transformation with phase $\pi/3$, one gets a $Z_6$
transformation which acts only on the $Q,\bar{Q}$ fields, and is just
the $Z_{2F}$ symmetry of Section~\ref{sec:indep} which was shown to be
contained in the continuous global symmetries as a discrete
subgroup. It seems that this theory does not have
interesting discrete symmetries. This conclusion is changed by the
presence of the quantum modified constraint. 

Let us, for example, examine the case when the operators $B_1,\bar{B}_1$
and $T$ acquire expectation values. In this case, the $SU(3)\times SU(3)$
non-Abelian global symmetries as well as $U(1)_R$ are left unbroken by
the VEV's, while both $U(1)_A$ and $U(1)_B$ are broken. 
However, one can combine a discrete $Z_6$ subgroup of $U(1)_B$ 
\begin{equation} B_1\to e^{2\pi i \frac{3}{6}} B_1, \; \; 
   \bar{B}_1\to e^{-2\pi i \frac{3}{6}} \bar{B}_1,\end{equation}
with the discrete $Z_4$ subgroup of $U(1)_A$ 
\begin{equation} B_1\to -B_1, \; \; \bar{B}_1\to -\bar{B}_1, \; \; T\to T,
\end{equation}
to find a $Z_{12}$ transformation which leaves $B_1,\bar{B}_1$ and $T$
invariant. Thus there is an unbroken $Z_{12}$ with charges
$q_{12}=12(Q_A/6+Q_B/4)$. These $Z_{12}$ charges are:
\begin{equation} 
 A: 3,\qquad Q:-1, \qquad \bar{Q}:-5,\qquad B_3:6,\qquad \bar{B}_3,M_0:-6,
\end{equation}
and all other fields have zero $Z_{12}$ charges. 
Because of the two constraints in the low-energy theory, one has to
exclude, for example, the fields $B_1$ and $B_3$ from the low-energy
spectrum, since the 
constraints give linear equations for them and
are hence not independent degrees of freedom in the low-energy theory.
Therefore these contribution of these fields to anomaly matching should not
be taken into account. The discrete
anomaly matching conditions are:

\begin{tabbing}
\hspace*{2cm}\= \hspace*{3cm} \= \hspace*{2.5cm} UV \hspace*{2cm} 
\= \hspace*{3cm}IR \hspace*{1cm} \\
\>  \begin{minipage}[c]{2cm} \[ SU(3)_Q^2Z_{12} \] \end{minipage} 
\> \begin{minipage}[c]{6cm} \[ -6  \] \end{minipage} 
\> \begin{minipage}[c]{6cm}
\[ -18\] \end{minipage} \\
\>  \begin{minipage}[c]{2cm} \[ SU(3)_{\bar{Q}}^2Z_{12} \] \end{minipage} 
\> \begin{minipage}[c]{6cm} \[ -30 \] \end{minipage} 
\> \begin{minipage}[c]{6cm}
\[  -18 \]
\end{minipage} \\
\>  \begin{minipage}[c]{2cm} \[ Z_{12}(\mbox{gravity})^{2} \] \end{minipage} 
\> \begin{minipage}[c]{6cm} \[ -48 \] \end{minipage} 
\> \begin{minipage}[c]{6cm}
\[  -60\]
\end{minipage} \\
\>  \begin{minipage}[c]{2cm} \[ Z_{12}^3\] \end{minipage} 
\> \begin{minipage}[c]{6cm} \[ -144 \times 12 \] \end{minipage} 
\> \begin{minipage}[c]{6cm}
\[ -180\times 12\]
\end{minipage} \\
\>  \begin{minipage}[c]{2cm} \[ U(1)_R^2 Z_{12}\] \end{minipage} 
\> \begin{minipage}[c]{6cm} \[  -48
 \] \end{minipage} 
\> \begin{minipage}[c]{6cm}
\[-60 \] 
\end{minipage} \\ \vspace*{-0.5cm}
\>  \begin{minipage}[c]{2cm} \[  U(1)_R Z_{12}^2\] \end{minipage}
\> \begin{minipage}[c]{6cm} \[ -54\times 12 \] \end{minipage} 
\> \begin{minipage}[c]{6cm} \[ -30\times 12 \]
\end{minipage} \\
\end{tabbing}
All anomaly matching conditions for $Z_{12}$ are satisfied.

Next we consider the case with no fundamentals, $e.g.$ $SU(6)$ with
$\Ythreea$. This theory has a discrete $Z_6$ symmetry. The low-energy
theory has two branches. On one branch there is a dynamically
generated superpotential. On the other branch there is a moduli space
of vacua described by the VEV of the operator $T=A^4$ and no 
superpotential:

\begin{equation} \begin{array}{c|c|cc}
& SU(6) & U(1)_R & Z_6 \\ \hline
A & \Ythreea & -1 & 1 \\ \hline \hline
T=A^4 & 1 & -4 & 4 \end{array} \end{equation}
This description matches the $U(1)_{R}(\mbox{gravity})^{2}$ and the 
$U(1)_{R}^{3}$ anomalies.
However, checking for example the $Z_{6}({\rm gravity})^2$ anomalies, 
we find that
they do not match ($Z_{6}({\rm gravity})^2$ is $20$ in the UV and $4$
in the IR,
and the difference is not divisible by $3$). We expect that, 
analogously to
the case of $SO(N)$ with $N-4$ vectors discussed in 
Section~\ref{sec:N-4}, the reason for the failure of
anomaly matching is the spontaneous breaking of the $Z_6$ discrete
symmetry. In the following, we show that this is indeed the case: $Z_6$
is spontaneously broken to $Z_2$. To see this, we start with the
$SU(6)$ theory with $\Ythreea +4(\Yfund +\overline{\Yfund})$. This
theory is s-confining, with the confining spectrum~\cite{sconf}
\begin{displaymath}
\label{su6four}
\begin{array}{c|c|ccccc} 
 & SU(6) & SU(4)& SU(4)& U(1)_B & U(1)_A & U(1)_R \\ \hline  
 A & \Ythreea & 1 & 1 & 0 & -4 & -1 \\
 Q & \Yfund & \Yfund & 1 & 1 & 3 & 1 \\
 \bar{Q} & \overline{\Yfund} & 1 & \Yfund & -1 & 3 & 1 \\ \hline \hline
M_0=Q\bar{Q}& & \Yfund & \Yfund & 0 & 6 & 2 \\
M_2=QA^2\bar{Q} & & \Yfund & \Yfund & 0 & -2 & 0 \\
B_1=AQ^3 & & \overline{\Yfund} & 1 & 3 & 5 & 2 \\
\bar{B}_1 =A \bar{Q}^3 & & 1 & \overline{\Yfund} & -3 & 5 & 2 \\
B_3=A^3 Q^3 & & \overline{\Yfund} & 1 & 3 & -3 & 0 \\
\bar{B}_3=A^3 \bar{Q}^3 & & 1 & \overline{\Yfund} & -3 & -3 & 0 \\
T=A^4 & & 1 & 1 & 0 & -16 & 4 \end{array}
\end{displaymath}
and a confining superpotential 
\begin{eqnarray}
\label{su6conf}
W_{dyn}&=&\frac{1}{\Lambda^{11}} \Big( M_0 B_1\bar{B}_1T +B_3\bar{B}_3 M_0
+ M_2^3M_0+TM_2M_0^3+ \nonumber \\
&& \bar{B}_1B_3M_2+B_1\bar{B}_3M_2 \Big). 
\end{eqnarray}
To obtain the $SU(6)$ theory with no flavors, we add a mass term
$m_{ij} M_0^{ij}$ with $\det m\neq 0$ to the superpotential in
Eq.~(\ref{su6conf}). One can see that the effect of the  $m
M_0$ term is to break the global $SU(4)\times SU(4)\times U(1)_B\times
U(1)_A \times U(1)_R$ symmetries to $U(1)_R \times Z_6$,  because
the quantum numbers of $m$ under these
global symmetries are
$(\overline{\Yfund},\overline{\Yfund},0,-6,0)$. Examination of the
solutions to the equations of motion obtained from the superpotential
of (\ref{su6conf}) with the $mM_0$ mass term shows that there is
a branch of solutions with $\langle M_0 \rangle = \langle B_1 \rangle
= \langle \bar{B}_1 \rangle = \langle B_3 \rangle =
 \langle \bar{B}_3 \rangle =0$, $\langle T \rangle$ arbitrary and 
$M_2^3=\Lambda^{11} m$. This solution with arbitrary value of $\langle
T \rangle$ is the branch with the moduli space of vacua. We can see
that this branch is characterized by a VEV for the operator $M_2$,
which carries $Z_6$ charge two, and hence $Z_6$ is spontaneously broken
to $Z_2$. One can easily check that the discrete anomaly matching
conditions involving the $Z_2$ are all satisfied. 

The order parameter $M_{2}$ of $Z_{6} \rightarrow Z_{2}$ breaking 
involves extra flavors $Q$, $\bar{Q}$, which do not exist in 
the SU(6) theory with a three-index anti-symmetric tensor and no flavors 
of our interest.  However, the 
expectation value of $M_2$ corresponds to that of the
$A^2W_{\alpha}W^{\alpha}$ operator which does not involve heavy flavors.  
The chiral anomaly equation in the $SU(6)$ theory 
with the four massive flavors of mass $m$ implies 
that~\cite{KS}\footnote{This is a supersymmetric 
generalization of the anomaly equation for gauge-covariant axial 
currents $\bar{\psi} T^{a} \gamma^{\mu} \gamma_{5} \psi$.  A term in the 
anomaly equation with supercovariant derivatives $\bar{D}^{2}$ can 
be dropped because of the translational invariance and supersymmetry 
of the vacuum.}
\begin{equation}
m \langle 
\bar{Q}T^aQ \;{\rm Tr}\; T^a A^2 \rangle 
=-\frac{1}{32\pi^2} \langle {\rm Tr}\; T^a W_{\alpha}W^{\alpha} \;
{\rm Tr}\; T^a A^2 \rangle = \Lambda^{11/3} m^{4/3} = \Lambda_{LE}^5,
\end{equation}
where $T^a$'s are the $SU(6)$ generators, $W_{\alpha}$ the field-strength
chiral superfield, and $\Lambda_{LE}$ is the dynamical $SU(6)$ scale
after we integrate out the four flavors $Q$, $\bar{Q}$. The $SU(6)$
group is broken to a pure $SU(3)\times SU(3)$ 
theory if $A$ is given an expectation value. The glueball field of the
pure $SU(3)$
theory can be identified with $A^2W_{\alpha}W^{\alpha}$. The
expectation value for $M_2$ signals that the field 
$A^2W_{\alpha}W^{\alpha}$ has a non-vanishing VEV, that is gaugino
condensation in the $SU(3)\times SU(3)$ theory.

\subsection{The ISS Model}
Our next example is the ISS model: $SU(2)$ with a three-index
symmetric tensor $t$. It has been conjectured in Ref.~\cite{ISS} that
this theory confines without generating a confining
superpotential. The basis of this conjecture is that the single
independent gauge invariant $X=t^4$ itself satisfies the 't Hooft
anomaly matching conditions for the $U(1)_R$ which is the only
continuous symmetry of the theory. However, there is also a discrete
global symmetry in this theory. The symmetries are
\begin{equation} \begin{array}{c|c|cc}
& SU(2)& U(1)_R & Z_{10} \\ \hline
t & \Ythrees & -\frac{1}{5} & 1 \\ \hline \hline
X=t^4 & 1 & -\frac{4}{5} & 4 \end{array}
\end{equation}
The discrete anomaly matching conditions are:

\begin{tabbing}
\hspace*{2cm}\= \hspace*{3cm} \= \hspace*{2.5cm} UV \hspace*{2cm} 
\= \hspace*{3cm}IR \hspace*{1cm} \\
\>  \begin{minipage}[c]{2cm} \[ Z_{10}(\mbox{gravity})^{2} \] \end{minipage} 
\> \begin{minipage}[c]{6cm} \[ 4 \] \end{minipage} 
\> \begin{minipage}[c]{6cm}
\[  4 \]
\end{minipage} \\
\>  \begin{minipage}[c]{2cm} \[ Z_{10}^3\] \end{minipage} 
\> \begin{minipage}[c]{6cm} \[ 4\] \end{minipage} 
\> \begin{minipage}[c]{6cm}
\[ 64  \]
\end{minipage} \\
\>  \begin{minipage}[c]{2cm} \[ U(1)_R^2 Z_{10}\] \end{minipage} 
\> \begin{minipage}[c]{6cm} \[  144
 \] \end{minipage} 
\> \begin{minipage}[c]{6cm}
\[ 324 \] 
\end{minipage} \\
\>  \begin{minipage}[c]{2cm} \[  U(1)_R Z_{10}^2\] \end{minipage}
\> \begin{minipage}[c]{6cm} \[ -24 \] \end{minipage} 
\> \begin{minipage}[c]{6cm} \[ -144 \]
\end{minipage} \\
\end{tabbing}
All discrete $Z_{10}$ anomaly matching conditions are satisfied mod
$10$. This seems to be a strong argument in favor of this theory. This
is however not the case, because we will argue that most of the
anomaly matching constraints follow from the anomaly matching for the
$U(1)_R$. The reason is that one can combine the $Z_{10}$
transformation with a discrete $U(1)_R$ transformation to get a $Z_2$
$R$-symmetry which together with the $U(1)_R$ symmetry is equivalent to
the $U(1)_R\times Z_{10}$. Thus the non-trivial part is only a $Z_2$,
under which the fermionic component of $t,X$ and the $SU(2)$ gauginos switch
sign. However, in the case of a $Z_2$ symmetry, 
neither the $Z_2({\rm gravity})^2$
nor the $Z_2^3$ anomalies yield any anomaly matching constraints 
because of the possible $N/2=1$ and $N^{3}/8=1$ terms in the matching 
equations.
The only non-trivial piece of information is the correlation between
the $Z_2({\rm gravity})^2$ and the $Z_2^3$ anomalies. If there is a contribution
from a massive Majorana fermion with charge $1$ to the $Z_2({\rm gravity})^2$
anomalies, there must a contribution to the $Z_2^3$ anomalies as well. 
Thus assuming that charge fractionalization does not occur in the 
heavy spectrum, either both
$Z_2({\rm gravity})^2$ and $Z_2^3$ have to match or neither of them. 
But even this
correlation is trivial, since in a $Z_2$ symmetry one can have only
charges $0$ or $1$, which have the same contribution to 
$Z_2({\rm gravity})^2$ and
$Z_2^3$. Thus one does not gain any information whatsoever from these
two anomalies. The only non-trivial ones are the $U(1)_R^2Z_2$ and
the $U(1)_RZ_2^2$, both of which are Type II and thus even if they
do not match we could not completely exclude the conjectured
spectrum. These anomalies are:
\begin{tabbing} \hspace*{2cm}
\= \hspace*{3cm} \= \hspace*{2.5cm} UV \hspace*{2cm} 
\= \hspace*{3cm}IR \hspace*{1cm} \\

\>  \begin{minipage}[c]{2cm} \[ U(1)_R^2 Z_{2}\] \end{minipage} 
\> \begin{minipage}[c]{6cm} \[  -37
 \] \end{minipage} 
\> \begin{minipage}[c]{6cm}
\[ -81 \] 
\end{minipage} \\

\>  \begin{minipage}[c]{2cm} \[  U(1)_R Z_{2}^2\] \end{minipage}
\> \begin{minipage}[c]{6cm} \[ -7 \] \end{minipage} 
\> \begin{minipage}[c]{6cm} \[  -9 \]
\end{minipage} \\
\end{tabbing}
Both anomalies are matched mod $2$. We conclude that the ISS model can
not be excluded using discrete anomaly matching, which gives a weak
additional evidence for the conjecture of Ref.~\cite{ISS}.

\subsection{Kutasov-type Duality}

Our final $N=1$ supersymmetric example in this section is the
Kutasov-type~\cite{Kutasov,otherKutasov} dual of $SO(N)$ with a traceless
symmetric tensor
and $F$ vectors and a tree-level superpotential for the symmetric
tensor. This theory  has been first studied by
Intriligator~\cite{Intrdual}. The reason we chose this theory is that
it has two separate discrete symmetries. One discrete symmetry arises
from the presence of the tree-level superpotential which explicitly
breaks a global $U(1)$ symmetry to its discrete subgroup. The other
source for a discrete symmetry is that we have an $SO$ theory with
vectors and thus there is a discrete $Z_{2F}$ symmetry present. This
theory will be an example for extremely non-trivial matching of discrete 
anomalies, including even mixed discrete anomalies.

The field content and the symmetries of the electric theory are:
\begin{equation}
\begin{array}{c|c|cccc}
& SO(N) & SU(F) & U(1)_R & Z_{2F} & Z_{(k+1)F} \\ \hline
X & \Ysymm & 1 & \frac{2}{k+1} & 0 & F \\
Q & \Yfund & \Yfund & 1-\frac{2(N-2k)}{(k+1)F} & 1 & -(N+2) 
\end{array}
\end{equation}
The superpotential of the electric theory is 
\begin{equation} W_{el}={\rm Tr}\, X^{k+1}.\end{equation}
The field content and the symmetries of the dual magnetic theory is
given by~\cite{Intrdual} 
\begin{equation} 
\begin{array}{c|c|cccc}
& SO(\tilde{N}) & SU(F) & U(1)_R & Z_{2F} & Z_{(k+1)F} \\ \hline
q & \Yfund & \overline{\Yfund}& 1-\frac{2(\tilde{N}-2k)}{(k+1)F}& -1 &
N+2-kF \\
Y & \Ysymm & 1 & \frac{2}{k+1} & 0 & F \\
M_j & 1 & \Ysymm & \frac{2(j+k)}{k+1}-4\frac{N-2k}{(k+1)F} & 2 &
(j-1)F -2(N+2) \end{array}\qquad, \end{equation}
where $\tilde{N}=k(F+4)-N$, 
$j=1,2,\ldots ,k$ and the superpotential of the magnetic theory is 
\begin{equation} W_{magn}={\rm Tr}\, Y^{k+1}+\sum_{j=1}^k M_jY^{k-j}q^2.
\end{equation}
The fields $M_{j}$ match the gauge-invariant polynomials 
$X^{j-1}Q^{2}$ in the electric theory.
The discrete charges in the magnetic theory have been assigned such
that the discrete symmetries are anomaly free, the magnetic superpotential
is invariant under the discrete symmetries and the gauge singlets
$M_j$ in the magnetic theory have the same charge as $X^{j-1}Q^2$ of the
electric theory. As described in Section~\ref{sec:discrete}, 
the electric $SO(N)$
theory also has a ${\cal P}$ outer automorphism if $N$ is
even. Furthermore, depending on whether $F$ and $(k+1)$ are even or
odd, some or all of the discrete symmetries may be contained in the
continuous global symmetries. Table~\ref{tab:map} shows that the
non-trivial discrete symmetries are always in one to one
correspondence between the electric and the magnetic 
theories, and
also gives the mapping of the discrete symmetries which is determined by
comparing the baryon type $B_p^{(n_1,\ldots ,n_k)}=W_{\alpha}^p 
Q_{(1)}^{n_1}\ldots Q_{(k)}^{n_k}$ operators with their magnetic
analog $Y^{(k-1)(k-p)} \tilde{B}_{\tilde{p}}^{(\tilde{n}_1,\ldots
,\tilde{n}_k)} $, where $Q_{(i)}=X^iQ$, $\tilde{p}=2k-p$, and
$\tilde{n}_{l}=F-n_{k+1-l}$ (for details of this
mapping, see Ref.~\cite{Intrdual}). The values of $N$, $F$ and
$(k+1)$ in the Table~\ref{tab:map} 
stand for $N$ mod $2$, $F$ mod $2$ and
$(k+1)$ mod $2$. In Table~\ref{tab:map} we have used that for $F$ odd
$Z_{2F}$ is equivalent to $Z_2\times Z_F$, where the $Z_F$ factor can
be identified with the center of $SU(F)$. Furthermore, if
$(k+1)$ and $F$ are relatively prime, 
$Z_{(k+1)F}$ is equivalent to $Z_{k+1}\times Z_{F}$, and the $Z_{F}$ 
factor can again be identified with the center of the $SU(F)$ symmetry.  Since 
$k$ and $F$ are arbitrary integers in this theory, however, we quote 
the discrete symmetries for generic $k$ and $F$ in Table~\ref{tab:map}.
\begin{table} 
\[
\begin{array}{|ccc|c|c|c|} \hline
N & F & (k+1)& {\rm Electric} &
{\rm Magnetic} & {\rm Mapping} \\ \hline
0 & 0 & 0 & {\cal P}\times Z_{2F}\times Z_{(k+1)F} & {\cal P}\times
  Z_{2F} \times Z_{(k+1)F}&  {\cal P}\leftrightarrow  {\cal P},  
Z_{2F}\leftrightarrow {\cal P}  Z_{2F}, \\
&&&&&Z_{(k+1)F}\leftrightarrow Z_{(k+1)F}\\ \hline 
0 & 0 & 1 & {\cal P}\times Z_{2F}\times Z_{(k+1)F}& {\cal P}\times
Z_{2F} \times Z_{(k+1)F} &  {\cal P}\leftrightarrow  {\cal P},  
Z_{2F}\leftrightarrow Z_{2F},
\\&&&&&Z_{(k+1)}\leftrightarrow Z_{(k+1)F}\\ \hline
0&1&0& {\cal P}\times Z_{(k+1)F} & Z_2 \times Z_{(k+1)F} &  
{\cal P}\leftrightarrow Z_2,  Z_{(k+1)F}\leftrightarrow  {\cal P}
Z_{(k+1)F} \\ \hline
0&1&1 &  {\cal P}\times  Z_{(k+1)F} &  {\cal P}\times  Z_{(k+1)F} &
 {\cal P}\leftrightarrow  {\cal P} ,  Z_{(k+1)F}\leftrightarrow 
Z_{(k+1)F} \\ \hline
1&0&0&  Z_{2F}\times Z_{(k+1)F} & Z_{2F}\times Z_{(k+1)F}&  
Z_{2F}\leftrightarrow
    Z_{2F}, \\ &&&&&Z_{(k+1)F}\leftrightarrow Z_{(k+1)F}\\ \hline 
1&0&1& Z_{2F}\times Z_{(k+1)F} & Z_{2F}\times Z_{(k+1)F}&  
Z_{2F}\leftrightarrow
    Z_{2F}, \\ &&&&&Z_{(k+1)F}\leftrightarrow Z_{(k+1)F}\\ \hline 
1&1&0& Z_2\times Z_{(k+1)F}& {\cal P}\times Z_{(k+1)F}&  
Z_2\leftrightarrow {\cal P}, Z_{(k+1)F}\leftrightarrow {\cal P} Z_{(k+1)F} \\ \hline
1&1&1&  Z_2\times Z_{(k+1)F}& Z_2\times Z_{(k+1)F}& Z_2\leftrightarrow Z_2,
Z_{(k+1)F}\leftrightarrow Z_{(k+1)F}\\ \hline\end{array}
\]
\caption{The mapping of discrete symmetries in the Kutasov-type duality
  of Ref.~\protect\cite{Intrdual} depending on the values of
$N,F$ and $k$.\label{tab:map}}
\end{table}
Note that in the case when $N$ is even and $F,(k+1)$ are odd (the
fifth row in Table~\ref{tab:map}), the mapping of the $Z_{2F}$
symmetries is non-trivial: the  generator $\omega$ is 
mapped to $-\omega$.

From the point of view of anomaly matching, we do not have to check
the discrete anomalies individually for every separate case. We will
check the anomaly matching conditions for the full $Z_{2F}\times
Z_{(k+1)F}$ symmetries, for any value of $N$, $F$ and $k$. In some
cases, part of these discrete symmetries is already contained in the
continuous global symmetries; anomaly matching for that piece should
follow from anomaly matching for the continuous symmetries, but it must
still be satisfied. Thus we do not loose any information by checking
anomaly matching for the bigger group. The effect of the mixing of the
discrete symmetries with the color-parity transformation 
can be taken into account
by adding $Z_{2F}$ charge $kF$ to every field carrying the first
$SO(\tilde{N})$ color and $Z_{(k+1)F}$ charge $k(k+1)F^2/2$ 
to the same fields.  The charge assignments for fermion fields
used for checking the 
anomaly matching are given in Table~\ref{tab:charges}.
\begin{table}
\[
\begin{array}{c|c|cccc}
& SO(N,\tilde{N}) & SU(F) & (k+1)F(R-1) & Z_{2F} & Z_{(k+1)F} \\ 
\hline
X & \frac{N(N+1)}{2}-1 & 1 & -(k-1)F & 0 & F \\
Q & N & \Yfund & -2(N-2k) & 1 & -(N+2) \\ \hline\hline
q^{1} & 1 & \overline{\Yfund}& -2(\tilde{N}-2k)& kF-1 &
N+2-kF+\frac{k(k+1)F^{2}}{2} \\
q^{i} & \tilde{N}-1 & \overline{\Yfund}& -2(\tilde{N}-2k)& -1 &
N+2-kF \\
Y^{1i} & \tilde{N}-1 & 1 & -(k-1)F & kF & F+\frac{k(k+1)F^{2}}{2} \\
Y^{ij} & \frac{\tilde{N}(\tilde{N}-1)}{2} & 1 & -(k-1)F & 0 & F \\
M_j & 1 & \Ysymm & F(2j+k-1)-4(N-2k) & 2 &
(j-1)F -2(N+2) \\
\tilde{\lambda}^{1i} & \tilde{N}-1 & 1 & (k+1)F & kF & \frac{k(k+1)F^{2}}{2}
\end{array} 
\]
\caption{The charge assignments
used for anomaly matching in the
  Kutasov-type $SO(N)$ duality with a symmetric tensor.
The $R$-charges are for the fermionic components and are normalized 
in such a way that all fields carry integer charges, and $\tilde{\lambda}$ is
the gaugino in the dual theory. 
\label{tab:charges}} 
\end{table}

Since the expressions for the anomalies are
sometimes quite lengthy, we quote only the simplified forms of
the differences between the anomalies of the magnetic and the electric
theories.  If a difference is given $\mbox{mod}\; 2F$ or mod $(k+1)F$, the 
expression is given after removing terms that are manifestly multiples
of $2F$ or $(k+1)F$ and thus are 
irrelevant to the anomaly matching conditions.

\begin{tabbing}
\= \hspace*{3cm} \= \hspace*{3cm} Anomalies$_{\it electric} -$ 
Anomalies$_{\it magnetic}$  \\ 
\>  \begin{minipage}[c]{2.5cm} \[ SU(F)^2Z_{2F}\] \end{minipage} 
\> \begin{minipage}[c]{12cm} \[-2Fk \] \end{minipage} \\

\>  \begin{minipage}[c]{2.5cm} \[ Z_{2F} (\mbox{gravity})^{2} \] \end{minipage} 
\> \begin{minipage}[c]{12cm} \[ (5-F)Fk -2F(4k^{2}+Fk^{2}-kN )
\] \end{minipage}\\ 
\end{tabbing}
One can see that the $ Z_{2F}(\mbox{gravity})^{2}$ anomaly matches mod $2F$ only
if $k$ is even or if $F$ is odd.  If $k$ is odd and $F$ is even,
there must be a massive Majorana fermion with $Z_{2F}$ charge
$F$. Since $2F$ is divisible by four in this case, we do not expect to
see the effect of this fermion in the $Z_{2F}^3$ anomaly. 
\vspace*{-1cm}
\begin{tabbing}
\= \hspace*{3cm} \= \hspace*{4cm} \\
\>  \begin{minipage}[c]{2.5cm} \[ Z_{2F}^3\] \end{minipage} 
\> \begin{minipage}[c]{12cm} \[ -F^3k^2(Fk-3) \; {\rm mod} \; 2F\] 
\end{minipage} \\ 
\end{tabbing}
This difference is as expected always a multiple of $2F$. This is
obvious if $F$ or $k$ is even. However, if both of them are odd then
$Fk-3$ is even and the above expression is again a multiple of $2F$.
\vspace*{-1cm}
\begin{tabbing}
\= \hspace*{3cm} \= \hspace*{4cm} \\
\>  \begin{minipage}[c]{2.5cm} \[ U(1)_R^2 Z_{2F}\] \end{minipage} 
\> \begin{minipage}[c]{12cm} \[ \frac{F^3(1+F)(1-k)k(1+k)}{3} 
\; {\rm mod} \; 2F
 \] \end{minipage} \\ \end{tabbing}
This is always a multiple of $2F$ since $(k-1)k(k+1)$ contains at
least one number divisible by three and one divisible by two.
\vspace*{-1cm}
\begin{tabbing}
\= \hspace*{3cm} \= \hspace*{4cm} \\
\>  \begin{minipage}[c]{2.5cm} \[ U(1)_R Z_{2F}^2\] \end{minipage}
\> \begin{minipage}[c]{12cm} \[ -4F^2k\Big[ 4k+Fk+Fk^2-2N\Big]
 \] \end{minipage} \\ \end{tabbing}
This is obviously a multiple of $2F$.
\vspace*{-1cm}
\begin{tabbing}
\= \hspace*{3cm} \= \hspace*{4cm} \\
\>  \begin{minipage}[c]{2.5cm} \[ SU(F)^2Z_{(k+1)F}\] \end{minipage} 
\> \begin{minipage}[c]{12cm} \[ 3Fk(1+k)
\] \end{minipage} \\ \end{tabbing}
This is obviously a multiple of $(k+1)F$.
\vspace*{-1cm}
\begin{tabbing}
\= \hspace*{3cm} \= \hspace*{4cm} \\
\>  \begin{minipage}[c]{3.5cm} \[ Z_{(k+1)F} (\mbox{gravity})^{2}\] \end{minipage} 
\> \begin{minipage}[c]{12cm} \[ -F(1+k)\frac{F(F-3)k}{4}\; {\rm mod} \;
  (k+1)F \] \end{minipage} \\ \end{tabbing}
If $k$ is even, then the anomalies are matched mod $(k+1)F$, since
$F(F-3)$ is always even. If $k$ is odd and $F$ is 0 or 3 mod 4,
then the anomalies are still matched mod $(k+1)F$. However if $k$ is
odd and $F$ is 1 or 2 mod 4, then the anomalies are matched
only mod $(k+1)F/2$, which signals the presence of odd number of
massive Majorana fermions with charge $(k+1)F/2$. We expect to see the
effect of these fermions in the $Z_{(k+1)F}^3$ anomaly for odd $k$
and $F=1$ mod 4; for odd $k$ and $F=2$ mod 4, $(k+1)F$ is divisible 
by four and hence $((k+1)F)^{3}/8$ term cannot be distinguished from 
mod $(k+1)F$ freedom.
\vspace*{-1cm}
\begin{tabbing}
\= \hspace*{3cm} \= \hspace*{4cm} \\
\>  \begin{minipage}[c]{2.5cm} \[ Z_{(k+1)F}^3\] \end{minipage} 
 \> \begin{minipage}[c]{12cm}
\begin{eqnarray} && \frac{{F^3}k ( 1 + k ) }{8}
       \Big[ 61Fk + 23{F^2}k - 31F{k^2} + 
         35{F^2}{k^2} - 2{F^3}{k^2} + {F^4}{k^2}  \nonumber \\ &&+ 
         4{F^3}{k^3} + 4{F^4}{k^3} + 
         14{F^3}{k^4} + 5{F^4}{k^4} + 
         2{F^4}{k^5} - 76N  \nonumber \\ && - 2{F^3}{k^2}N - 
         4{F^3}{k^3}N - 2{F^3}{k^4}N - 12{N^2}
          \Big]
\; {\rm mod} \; (k+1)F \nonumber \end{eqnarray} 
\end{minipage} \\ \end{tabbing}
One can show that this expression indeed satisfies all the
requirements.  The terms $-76N-12N^{2}$ give a 
multiple of 8 and can be dropped.  Then all terms in the square 
bracket have a factor of $F$, and hence the difference is obviously a multiple 
of $(k+1)F$ if $F$ is even.  If $k=2n$, the difference is 
simplified to $-\frac{1}{2}{F^4}{n^2}( 1 + 2n) ( 5 + 7F)( 1 + N )$ 
modulo $(k+1)F$ and 
we again obtain a multiple of $(k+1)F$ because $F(5+7F)$ is even.  The 
non-trivial case is when $F=2m-1$, $k=2n-1$.  Then the difference 
reduces to $m((k+1)F)^{3}/8$ modulo $(k+1)F$; thus as
expected, one can see the presence of the massive Majorana fermions
only for odd $k$ and $F=1$ mod 4 ({\it i.e.}\/, odd $m$). 
\vspace*{-1cm}
\begin{tabbing}
\= \hspace*{3cm} \= \hspace*{4cm} \\
\>  \begin{minipage}[c]{2.5cm} \[ U(1)_R^2 Z_{(k+1)F}\] \end{minipage} 
\> \begin{minipage}[c]{12cm} \begin{eqnarray} && 
-\frac{F^{2}(k+1)}{12}\Big[32Fk + 3{F^2}k + {F^3}k - 464F{k^3} - 
  123{F^2}{k^3} - {F^3}{k^3}\Big] \nonumber \\ &&
 \; {\rm mod} \; (k+1)F \nonumber \end{eqnarray}
\end{minipage} \\ \end{tabbing}
By adding multiples of $(k+1)F$, one can simplify the difference to 
the form $\frac{{F^3}}{12}(F-1)(F+4) (k -1) k (k + 1)^2$.
The last factor $(k-1)k(k+1)$ is a multiple of 6, and $(F-1)(F+4)$ is 
an even number.  Therefore, the difference vanishes mod $(k+1)F$.
\vspace*{-1cm}
\begin{tabbing}
\= \hspace*{3cm} \= \hspace*{4cm} \\
\>  \begin{minipage}[c]{2.5cm} \[ U(1)_R Z_{(k+1)F}^2\] \end{minipage}
\> \begin{minipage}[c]{12cm} \begin{eqnarray}
&&\frac{-{F^2}( 1 + k)}{12} 
       \Big[ 32Fk + 3{F^2}k + {F^3}k - 
         176F{k^3} - 123{F^2}{k^3} - {F^3}{k^3}
          \Big]   \nonumber \\ &&
 \nonumber \\ &&{\rm mod} \; (k+1)F \nonumber \end{eqnarray}
\end{minipage} \\ \end{tabbing}
One can further simplify the difference to the form $\frac{{F^3}}{12}(F-1)(F+4) 
(k -1) k (k + 1)^2$ by adding multiples of $(k+1)F$.
This result is the same as the $U(1)_R^2 Z_{(k+1)F}$ and hence the 
anomalies match here as well.

For the mixed discrete anomalies $Z_{(k+1)F}^{2} Z_{2F}$, $Z_{(k+1)F} 
Z_{2F}^{2}$, and $U(1)_{R}Z_{(k+1)F} Z_{2F}$, either the charges or 
the multiplicity for both electric and magnetic degrees of freedom 
have a factor of $F$.  If $k$ is even, the greatest common divisor of 
$(k+1)F$ and $2F$ is $F$, and the anomalies need to be matched only 
mod $F$; therefore their anomaly matching is trivial for even $k$.  If 
$k$ is odd, then $(k+1)F$ is divisible by $2F$, thus the greatest 
common divisor of $2F$ and $(k+1)F$ is $2F$.  Therefore we would like 
to show that the differences in anomalies are multiples of $2F$ for odd 
$k=2n-1$.

\vspace*{-1cm}
\begin{tabbing}
\= \hspace*{3cm} \= \hspace*{4cm} \\
\>  \begin{minipage}[c]{3cm} \[ Z_{(k+1)F}^2Z_{2F} \] \end{minipage} 
\> \begin{minipage}[c]{13cm}
\begin{eqnarray} && 
\frac{-{F^2}k }{12}
       \Big[ 14F + 2{F^2} - 30Fk + 30{F^2}k - 
         3{F^3}k - 44F{k^2} + 76{F^2}{k^2}  \nonumber \\ &&- 
         12{F^3}{k^2} + 3{F^4}{k^2} + 
         9{F^3}{k^3} + 12{F^4}{k^3} + 
         42{F^3}{k^4} + 15{F^4}{k^4} + 
         6{F^4}{k^5} \nonumber \\ && + 168N - 6{F^3}{k^2}N - 
         12{F^3}{k^3}N - 6{F^3}{k^4}N \Big] \; \mbox{mod} \; 2F
\nonumber \end{eqnarray} 
\end{minipage} \\ \end{tabbing}
Substituting
$k=2n-1$ and leaving out multiples of $2F$, the difference becomes 
$\frac{1}{3}{F^3}( 1 + F ) n (2n -1) ( 1 + 2n - 3{F^2} n) $.  The 
factor $F(1+F)$ is even, and hence the term $-3F^{2}n$ can be dropped 
modulo $2F$.  Then the last three factors give $n(2n-1)(2n+1)$ which is a 
multiple of 3, thus the difference vanishes modulo $2F$.
\vspace*{-1cm}
\begin{tabbing}
\= \hspace*{3cm} \= \hspace*{4cm} \\
\>  \begin{minipage}[c]{3cm} \[ Z_{(k+1)F}Z_{2F}^2 \] \end{minipage} 
\> \begin{minipage}[c]{13cm} \begin{eqnarray} && 
\frac{-{F^2}k}{2}
       \Big[ -22 - F - 6k + 7Fk - 2{F^2}k + 
         8F{k^2} - 4{F^2}{k^2} + {F^3}{k^2} \nonumber \\ && + 
         6{F^2}{k^3} + 3{F^3}{k^3} + 
         8{F^2}{k^4} + 2{F^3}{k^4} - 8N - 
         2{F^2}{k^2}N - 2{F^2}{k^3}N \Big] 
\nonumber \end{eqnarray}
\end{minipage} \\ \end{tabbing}
By substituting $k=2n-1$, the difference can be
simplified to $F^3(9 + {F^2}) n (2n -1) \; \mbox{mod} \; 2F$,  
and the factor $n(2n-1)$ is always even; thus the anomaly is matched 
modulo $2F$.
\vspace*{-1cm}
\begin{tabbing}
\= \hspace*{3cm} \= \hspace*{4cm} \\
\>  \begin{minipage}[c]{3cm} \[ U(1)_R Z_{(k+1)F}Z_{2F}\] \end{minipage} 
\> \begin{minipage}[c]{13cm} \begin{eqnarray}&&
\frac{{F^3}k}{6}\Big[ 1 + F + 66k + 3Fk + 
       3{F^2}k + 5{k^2} - 7F{k^2} \nonumber \\ && - 
       9{F^2}{k^2} - 3{F^3}{k^2} - 9F{k^3} - 
       3{F^2}{k^3} + 9{F^2}{k^4} + 3{F^3}{k^4} - 
       6N  \nonumber \\ && + 30kN - 3FkN + 3{F^2}kN + 
       3F{k^3}N - 3{F^2}{k^3}N \Big] 
\; {\rm mod} \; 2F \nonumber
 \end{eqnarray}  \end{minipage} \\
\end{tabbing}
Substituting $k=2n-1$ we get for the difference $\frac{-2{F^3} 
}{3}(1 + F) (n-1) n (2n -1)\;\mbox{mod}\; 2F$.  The last factor is a 
multiple of 3 and hence the $U(1)_R Z_{(k+1)F}Z_{2F}$ anomalies match
modulo $2F$ as well. 

Thus we have seen that this example with two different discrete 
symmetries have all anomalies matched between the electric and the 
magnetic theories in a highly non-trivial manner. Note that all Type
II anomaly matching conditions including the correlation between the
$Z_{N} ({\rm gravity})^2$ and $Z_N^3$ anomalies are satisfied as
well.

\section{Excluded Models\label{sec:exclude}}
\setcounter{equation}{0}
\setcounter{footnote}{0}

We have seen several examples of discrete anomaly matching in the
previous two sections. In this section we will show examples of
theories where the conjectured low-energy spectrum does not satisfy
the discrete anomaly matching conditions, which means that the given
spectrum can not be the correct low-energy solution of the theory.

In the first example, we will consider the recently suggested chirally
symmetric phase of $N=1$ supersymmetric pure Yang-Mills
theory~\cite{Shifman}.
We will show that the chirally symmetric vacuum described by the
natural variable of the Veneziano--Yankielowicz Lagrangian \cite{VY}
does not satisfy
the discrete anomaly matching conditions and thus can be
excluded.  However, the concept of a chirally symmetric phase can not
be completely excluded, since there may be another set of massless
states which does satisfy the anomaly matching conditions. The
next set of examples will deal with the non-supersymmetric confining
examples conjectured in the early 80's~\cite{Albright,othernonsusy}. 
We will show that almost all
examples in this category which have a non-trivial discrete symmetry
can be excluded based on discrete anomaly matching. Finally, we
consider the self-dual $N=1$ supersymmetric theories based on
exceptional and orthogonal groups~\cite{Ramond,Distler,Karch,ourself}. 
We show that the discrete symmetries of the electric and the magnetic
theories can not be mapped to each other in the examples based on exceptional 
groups~\cite{Ramond,Distler,Karch}. However, since both the
electric and the magnetic theories are strongly coupled, one can not
exclude the presence of accidental symmetries. Thus this category of
theories can not be completely excluded based on discrete anomaly
matching, but the evidence for duality is made much weaker than it is in
other theories. The self-dual theories based on orthogonal
groups~\cite{ourself,Karch} do satisfy discrete anomaly matching once
the maximal number of meson fields is included in the dual theory as
elementary fields. 

\subsection{$N=1$ Supersymmetric Pure Yang-Mills Theories}

These theories do not have any continuous global symmetries. Their only
symmetry is a discrete $Z_{\mu (G)}$
$R$-symmetry, where $\mu (G)$ is the Dynkin index of the adjoint
representation of the gauge group $G$ (twice the dual Coxeter number
$h^{\vee}$).  
The $\lambda_{\alpha}$ gaugino (which is the only fermion in
these theories) carries one unit of the  discrete $Z_{\mu (G)}$
charge. 

The canonical description of the low-energy dynamics~\cite{pureYM} of
this theory is that gaugino condensation occurs, 
\begin{equation} 
\langle \lambda_{\alpha}\lambda^{\alpha}\rangle
=\omega_i\Lambda_G^3,\; \; \; i=1,2,\ldots ,\mu (G)/2,
\end{equation}
where the $\omega_i$'s are the $\mu (G)/2$ roots of unity. This
gaugino condensate 
breaks the discrete $Z_{\mu (G)}$ spontaneously to $Z_2$, and
the fields from the vector multiplet $W_{\alpha}$ form massive bound
states. The theory confines with chiral symmetry breaking. One does
not get any useful information from a $Z_2$ discrete symmetry, since
the massive Majorana fermions can modify both the $Z_2({\rm
  gravity})^2$ 
and the
$Z_2^3$ anomalies by one. 

However, it has been recently suggested by Kovner and 
Shifman~\cite{Shifman} that there might be another branch of the
theory on which spontaneous breaking of  $Z_{\mu (G)}$ does not occur,
but there are massless fermions at the origin. This
conclusion has been made in Ref.~\cite{Shifman} by examining the
vacuum structure of a modified Veneziano--Yankielowicz (VY)  Lagrangian~\cite{VY}
which is $Z_{\mu (G)}$ symmetric and reproduces all the Green's
functions for the fields of $ W_{\alpha}W^{\alpha}$. 
The modified VY Lagrangian suggests that there 
is a single massless fermion $\Phi =(W_{\alpha}
W^{\alpha})^{\frac{1}{3}}$ present in the low-energy theory.  If there
is indeed such a phase of the theory where $Z_{\mu (G)}$ is not
spontaneously broken, the discrete anomaly matching conditions
must be satisfied.  In the following we show that the discrete anomaly
matching conditions are satisfied neither with the
field $\Phi$ nor the field $S= W_{\alpha}W^{\alpha}$ as the only
massless composite field in the low-energy theory.

First we assume that the only
massless field is $\Phi =(W_{\alpha}W^{\alpha})^{\frac{1}{3}}$ as
suggested by the modified VY Lagrangian.
In this case, the $R$-charge of the fermionic component of $\Phi$ is
$-\frac{1}{3}$, which signals the fractionalization of the $Z_{\mu
  (G)}$ charges. Therefore, it is convenient to rescale the discrete
charges such that the gaugino of the high-energy theory has charge
$3$, and check the anomaly matching conditions for the resulting 
$Z_{3\mu (G)}$ symmetry.  Values of $\mu(G)$ for semi-simple gauge 
groups are listed in Table~\ref{algebras}.
\begin{table}
\begin{center}
\begin{tabular}{ccccl}
algebra & group & dim & $\mu (G)$ & $H^{*}(G; {\bf R})$ generators\\
\hline
$A_{n}$ & $SU(n+1)$ & $n(n+2)$ & $2(n+1)$ & 3, 5, $\cdots$, $2n+1$\\
$B_{n}$ & $SO(2n+1)$ & $n(2n+1)$ & $2(2n-1)$ & 3, 7, $\cdots$, $4n-1$\\
$C_{n}$ & $Sp(2n)$ & $n(2n+1)$ & $2(n+1)$ & 3, 7, $\cdots$, $4n-1$\\
$D_{n}$ & $SO(2n)$ & $n(2n-1)$ & $2(2n-2)$ & 3, 7, $\cdots$, $4n-5$, $2n-1$\\
$E_{6}$ & $E_{6}$ & 78 & 24& 3, 9, 11, 15, 17, 23\\
$E_{7}$ & $E_{7}$ & 133 & 36 & 3, 11, 15, 19, 23, 27, 35\\
$E_{8}$ & $E_{8}$ & 248 & 60 & 3, 15, 23, 27, 35, 39, 47, 59\\
$F_{4}$ & $F_{4}$ & 52 & 18 & 3, 11, 15, 23\\
$G_{2}$ & $G_{2}$ & 14 & 8 & 3, 11
\end{tabular}
\end{center}
\caption{\label{algebras}Dimensions, Dynkin index of adjoint 
representations for semi-simple Lie algebras, and the degrees of forms 
which generate the cohomology ring of the group manifold.}
\end{table}

The discrete anomalies for $G=SU(N)$ are ($\mu (G)=2N$):

\begin{tabbing} 
\hspace*{1.5cm} \= \hspace*{3cm} \= \hspace*{2.5cm} UV \hspace*{2cm} 
\= \hspace*{3cm}IR \hspace*{1cm} \\
\>  \begin{minipage}[c]{2.5cm} \[ Z_{6N} ({\rm gravity})^2 \] \end{minipage} 
\> \begin{minipage}[c]{6cm} \[ 3(N^2-1)\] \end{minipage} 
\> \begin{minipage}[c]{6cm} \[ -1 \]
\end{minipage} \\

\>  \begin{minipage}[c]{2.5cm} \[ Z_{6N}^3\] \end{minipage} 
\> \begin{minipage}[c]{6cm} \[ 27(N^2-1)\] \end{minipage} 
\> \begin{minipage}[c]{6cm} \[ -1 \]
\end{minipage} \\
\end{tabbing}

The difference in the $Z_{6N} ({\rm gravity})^2$ anomalies of the UV and the IR
descriptions is 2 mod $3N$, which means that the discrete anomalies can
not be matched for any value of $N$.  Recall that the $Z_{6N} ({\rm
  gravity})^2$ anomaly is Type I and 
must be matched irrespective of charge
fractionalization.  Therefore, this low-energy description of the pure
$SU(N)$ YM theories is excluded.

Next we consider the case of $SO(N)$ groups ($\mu (G)=2N-4$). The
discrete anomalies are:

\begin{tabbing} 
\hspace*{1.5cm} \= \hspace*{3cm} \= \hspace*{2.5cm} UV \hspace*{2cm} 
\= \hspace*{3cm}IR \hspace*{1cm} \\
\>  \begin{minipage}[c]{3cm} \[ Z_{6N-12} ({\rm gravity})^2 \] \end{minipage} 
\> \begin{minipage}[c]{6cm} \[ 3\frac{N(N-1)}{2}\] \end{minipage} 
\> \begin{minipage}[c]{6cm} \[ -1 \]
\end{minipage} \\

\>  \begin{minipage}[c]{2.5cm} \[ Z_{6N-12}^3\] \end{minipage} 
\> \begin{minipage}[c]{6cm} \[ 27\frac{N(N-1)}{2}\] \end{minipage} 
\> \begin{minipage}[c]{6cm} \[ -1 \]
\end{minipage} \\
\end{tabbing}

The difference in the $Z_{6N-12} ({\rm gravity})^2$ anomalies of the
UV and the IR 
descriptions is $3\frac{N(N-1)}{2}+1$ which should be divisible at
least by $3(N-2)$. However, $3N^2-3N+2$ is never divisible by $3N-6$,
thus we conclude that anomaly matching is not satisfied for
$SO(N)$ theories either. 

For $Sp(2N)$ groups $\mu (G)=2N+2$. The anomaly matching conditions
are: 
\begin{tabbing} 
\hspace*{1.5cm} \= \hspace*{3cm} \= \hspace*{2.5cm} UV \hspace*{2cm} 
\= \hspace*{3cm}IR \hspace*{1cm} \\
\>  \begin{minipage}[c]{3cm} \[ Z_{6N+6} ({\rm gravity})^2 \] \end{minipage} 
\> \begin{minipage}[c]{6cm} \[ 3N(2N+1)\] \end{minipage} 
\> \begin{minipage}[c]{6cm} \[ -1 \]
\end{minipage} \\

\>  \begin{minipage}[c]{2.5cm} \[ Z_{6N+6}^3\] \end{minipage} 
\> \begin{minipage}[c]{6cm} \[ 27N(2N+1)\] \end{minipage} 
\> \begin{minipage}[c]{6cm} \[ -1 \]
\end{minipage} \\
\end{tabbing}
The difference in the $Z_{6N+6} ({\rm gravity})^2$ anomalies is
$(N+1)(6N-3)+4$, which is never 
divisible by $3(N+1)$. Thus
the discrete anomaly matching constraints are not satisfied
for the $Sp(2N)$ groups either.  Furthermore, we have checked that none
of the similarly constructed solutions for the exceptional groups
$G_2,F_4,E_{6,7,8}$ satisfy the discrete anomaly matching
conditions.  Note that the
$Z_{3\mu (G)}({\rm gravity})^2$ anomalies are Type I and should be
satisfied under all circumstances.  We conclude that the most natural
implementation of a chirally symmetric vacuum of $N=1$ pure Yang-Mills
theories can be excluded based on discrete anomalies. 

However, this does not completely exclude
the idea of a chirally symmetric phase of $N=1$ pure
Yang-Mills theories. It excludes only a specific realization of it
described above. One
could, for example, try to match anomalies with the operator $S= 
W_{\alpha}W^{\alpha}$ instead of $\Phi$. 
Here no charge fractionalization occurs, and hence anomalies should be matched
mod $\mu (G)$.

The anomalies for $SU(N)$ are

\begin{tabbing} 
\hspace*{1.5cm} \= \hspace*{3cm} \= \hspace*{2.5cm} UV \hspace*{2cm} 
\= \hspace*{3cm}IR \hspace*{1cm} \\
\>  \begin{minipage}[c]{2.5cm} \[ Z_{2N} ({\rm gravity})^2 \] \end{minipage} 
\> \begin{minipage}[c]{6cm} \[ N^2-1\] \end{minipage} 
\> \begin{minipage}[c]{6cm} \[ 1 \]
\end{minipage} \\

\>  \begin{minipage}[c]{2.5cm} \[ Z_{2N}^3\] \end{minipage} 
\> \begin{minipage}[c]{6cm} \[ N^2-1\] \end{minipage} 
\> \begin{minipage}[c]{6cm} \[ 1 \]
\end{minipage} \\
\end{tabbing}

The differences in the anomalies are both $N^2-2$, which is divisible by $N$
only for $N=1,2$. Performing a similar analysis we find that the field
$S$ matches the discrete anomalies for $SO(N)$ only if $N$ is odd, while
it matches always for $Sp(2N)$.  None of the discrete anomalies for the
exceptional groups are matched by $S$.  Even
though anomalies are matched for some special cases by $S$,
generically it does not match the discrete anomalies and therefore 
we conclude that it
is not a likely candidate for a low-energy solution. 

As emphasized above, the failure of anomaly matching excludes only a
particular low-energy spectrum. It is in fact possible to find a set
of discrete $R$-charges that 
match the anomalies.  However, we cannot identify natural 
interpolating fields as composite operators of the field strength 
superfield.  The following construction is an example for a set of
$R$-charges which satisfy discrete anomaly matching. This construction
works for  all semi-simple gauge groups.  

As clear from the previous discussions, we would like to match
$\mbox{Tr} R = \mbox{dim}(G)$ modulo $\mu (G)/2$ and $\mbox{Tr} R^{3} = 
\mbox{dim}(G)$ modulo $\mu (G)$.  
A set of useful numbers for semi-simple Lie algebras is given in
Table~\ref{algebras}.
The last column in Table~\ref{algebras} 
shows the degrees $k$ of the forms which generate the cohomology 
ring on group manifolds;\footnote{They coincide with $2 e_{i} + 1$ 
where the $e_{i}$'s  are the exponents of the Lie algebra.}
they are $k$-forms which can be written as 
$\mbox{Tr} (g^{-1} d g)^{k}$ with group elements $g \in G$.  All other
elements of the cohomology ring   
are given by products of the generators (note that one cannot use the 
same generator more than once because they are all forms of odd 
degrees) and their linear combinations.  
In particular, the volume form is given by the product of 
all generators and hence the sum of the numbers in the last column 
must give the dimensions of the groups; this can be checked easily.  
Therefore, if one has a set of fermions whose $R$-charges are given by 
the degrees of cohomology generators, the $Z_{\mu(G)}(\mbox{gravity})^2$
anomalies are matched exactly.

Curiously enough, the $Z^3_{\mu (G)}$ anomalies are also matched modulo 
$\mu(G)$ with this set of $R$-charges.  The problem is to find 
interpolating fields for such $R$-charges using gauge invariant composite 
operators of field strength superfield $W_{\alpha}^{a}$ with spin $1/2$.
The operators 
$\omega_k^{a_1,\ldots ,a_k}\lambda_{\alpha_1}^{a_1}\ldots
\lambda_{\alpha_k}^{a_k}$, where the $\omega_{k}$ is the cohomology
generator of degree $k$ 
and $\lambda$'s are the gauginos, have the correct $R$-charges,
but the spinor indices $\alpha_i$ are totally symmetric for these 
operators and they cannot produce spin $1/2$ fermions.  Since massless
fields with higher 
spin cannot have consistent interactions, we exclude this 
choice of operators.  If there are operators which match the 
required $R$-charges, they necessarily need to involve derivatives
and hence are bound states with higher relative orbital angular momenta.  We 
find such composite spectrum to be highly unlikely.

\subsection{Non-supersymmetric Theories}

In this section, we examine several non-supersymmetric theories which
were conjectured to be confining based on the 't Hooft anomaly
matching conditions in the early 80's~\cite{Albright,othernonsusy}. 
We show that
most of the examples which have a non-trivial discrete symmetry do
not satisfy the discrete anomaly matching conditions and thus one can
exclude these conjectured spectra. We briefly comment on the recently
proposed duality for non-supersymmetric QCD \cite{John} at the end of
the section. 

The first example we consider is based on a non-supersymmetric $SU(4)$
theory with two massless left-handed Weyl fermions in the
antisymmetric tensor representation of $SU(4)$, and one in the adjoint
representation~\cite{Albright}. 
This theory was conjectured to be confining. The
global symmetries and the conjectured confining spectrum is given in
the table below.

\begin{equation} \begin{array}{c|c|ccc}
& SU(4)& SU(2) & U(1) & Z_{12} \\ \hline
A & \Yasymm & \Yfund & 2 & 1 \\
X & \Yoneoone & 1 & -1 & 1 \\ \hline \hline
(A^2X) & 1 & \Ysymm & 3 & 3 \end{array}
\end{equation}
All the continuous global anomalies ($SU(2)^2U(1)$,
$U(1)({\rm gravity})^2$ and $U(1)^3$) are matched between the
high-energy and the
confining spectrum. The discrete anomalies are:

\begin{tabbing} \hspace*{2cm}
\= \hspace*{3cm} \= \hspace*{2.5cm} UV \hspace*{2cm} 
\= \hspace*{2.75cm}IR \hspace*{1cm} \\
\>  \begin{minipage}[c]{2.5cm} \[ SU(2)^2Z_{12} \] \end{minipage} 
\> \begin{minipage}[c]{6cm} \[ 6  \] \end{minipage} 
\> \begin{minipage}[c]{6cm}
\[ 12 \] \end{minipage} \\

\>  \begin{minipage}[c]{2.5cm} \[ Z_{12} ({\rm gravity})^2 \] \end{minipage} 
\> \begin{minipage}[c]{6cm} \[ 27 \] \end{minipage} 
\> \begin{minipage}[c]{6cm}
\[  9 \]
\end{minipage} \\

\>  \begin{minipage}[c]{2.5cm} \[ Z_{12}^3\] \end{minipage} 
\> \begin{minipage}[c]{6cm} \[ 27 \] \end{minipage} 
\> \begin{minipage}[c]{6cm}

\[ 81\]
\end{minipage} \\

\>  \begin{minipage}[c]{2.5cm} \[ U(1)^2Z_{12}\] \end{minipage} 
\> \begin{minipage}[c]{6cm} \[  63
 \] \end{minipage} 
\> \begin{minipage}[c]{6cm}
\[ 81 \] 
\end{minipage} \\

\>  \begin{minipage}[c]{2.5cm} \[  U(1) Z_{12}^2\] \end{minipage}
\> \begin{minipage}[c]{6cm} \[ 9  \] \end{minipage} 
\> \begin{minipage}[c]{6cm} \[ 81 \]
\end{minipage} \\
\end{tabbing}

The $ U(1)^2Z_{12}$ anomaly matching is satisfied mod $12$ and
the $ Z_{12} ({\rm gravity})^2$ anomaly matching is 
satisfied mod $6$. However, while the $SU(2)^2Z_{12}$, the $
U(1)^2Z_{12}$ and the $Z_{12}^3$ 
anomalies must match mod $12$, they match only
mod $6$,  and hence the discrete anomaly matching conditions are
violated. In the absence of any dynamical explanation of spontaneous
breaking of $Z_{12}$, and since $SU(2)^2Z_{12}$ is a Type I anomaly,
 one has to consider this model excluded  based on
discrete anomaly matching. 

The next example is an $SU(5)$ theory with the field content
and the conjectured confining spectrum to be~\cite{Albright}

\begin{equation} \begin{array}{c|c|ccc}
& SU(5)& SU(6) & U(1) & Z_{42} \\ \hline
A & \Yasymm & \Yfund & 4 & 1 \\
Y & \overline{\Yoneoone} & 1 & -3 & 1 \\ \hline \hline
(A^2Y) & 1 & \Ysymm & 5 & 3 \end{array} \end{equation}

The anomalies with respect to the continuous flavor symmetries
are all matched, and so are all discrete anomalies except the
$SU(6)^2Z_{42}$ whose value is $10$ in the UV and $24$ in the IR, and
thus the difference is $14$. Since $SU(6)^2Z_{42}$ is a Type I anomaly,
this example is excluded as well.

Finally, we present two examples where all continuous anomalies are
matched but almost all of the discrete anomaly matching conditions are
violated. The first example is an $SU(9)$ gauge theory with
massless fermions and the confining spectrum:
\begin{equation} \begin{array}{c|c|ccc}
& SU(9) & U(1) & Z_{42} \\ \hline
A & \Yasymm & 5 & 1 \\
B & \overline{\Yfoura} & -1 & 1 \\ \hline \hline
6\times (A^2B)& 1 & 9 & 3 \end{array}  \end{equation}
The conjectured spectrum contains six different copies of the $(A^2B)$
bound state. The discrete anomalies are:

\begin{tabbing} \hspace*{2cm}
\= \hspace*{3cm} \= \hspace*{2.5cm} UV \hspace*{2cm} 
\= \hspace*{2.75cm}IR \hspace*{1cm} \\

\>  \begin{minipage}[c]{2.5cm} \[ Z_{42} ({\rm gravity})^2 \] \end{minipage} 
\> \begin{minipage}[c]{6cm} \[ 162 \] \end{minipage} 
\> \begin{minipage}[c]{6cm}
\[  18 \]
\end{minipage} \\

\>  \begin{minipage}[c]{2.5cm} \[ Z_{42}^3\] \end{minipage} 
\> \begin{minipage}[c]{6cm} \[ 162 \] \end{minipage} 
\> \begin{minipage}[c]{6cm}
\[ 162 \]
\end{minipage} \\

\>  \begin{minipage}[c]{2.5cm} \[ U(1)^2Z_{42}\] \end{minipage} 
\> \begin{minipage}[c]{6cm} \[  1026
 \] \end{minipage} 
\> \begin{minipage}[c]{6cm}
\[ 1458 \] 
\end{minipage} \\

\>  \begin{minipage}[c]{2.5cm} \[  U(1) Z_{42}^2\] \end{minipage}
\> \begin{minipage}[c]{6cm} \[ 54  \] \end{minipage} 
\> \begin{minipage}[c]{6cm} \[ 486 \]
\end{minipage} \\
\end{tabbing}

One can see that none of the discrete anomaly matching conditions
(except the  $ Z_{42}^3$) are satisfied, thus this example is excluded
as well.

Finally, we consider a theory based on an $SU(3)$ gauge group:
\begin{equation} \begin{array}{c|c|cccc}
& SU(3) & SU(2) & U(1) & Z_{30} \\ \hline
S & \Ysymm & \Yfund & 2 & 1 \\
W & \overline{\Ytwoone} & 1 & -1 & 1 \\ \hline \hline 
(S^2W) & 1 & \Ysymm & 3 & 3 \end{array}  \end{equation}

The discrete anomalies are:

\begin{tabbing} \hspace*{2cm}
\= \hspace*{3cm} \= \hspace*{2.5cm} UV \hspace*{2cm} 
\= \hspace*{2.75cm}IR \hspace*{1cm} \\
\>  \begin{minipage}[c]{2.5cm} \[ SU(2)^2Z_{30} \] \end{minipage} 
\> \begin{minipage}[c]{6cm} \[ 6  \] \end{minipage} 
\> \begin{minipage}[c]{6cm}
\[ 12 \] \end{minipage} \\

\>  \begin{minipage}[c]{2.5cm} \[ Z_{30} ({\rm gravity})^2 \] \end{minipage} 
\> \begin{minipage}[c]{6cm} \[ 27 \] \end{minipage} 
\> \begin{minipage}[c]{6cm}
\[  9 \]
\end{minipage} \\

\>  \begin{minipage}[c]{2.5cm} \[ Z_{30}^3\] \end{minipage} 
\> \begin{minipage}[c]{6cm} \[ 27 \] \end{minipage} 
\> \begin{minipage}[c]{6cm}
\[ 81\]
\end{minipage} \\

\>  \begin{minipage}[c]{2.5cm} \[ U(1)^2Z_{30}\] \end{minipage} 
\> \begin{minipage}[c]{6cm} \[  63
 \] \end{minipage} 
\> \begin{minipage}[c]{6cm}
\[ 81 \] 
\end{minipage} \\

\>  \begin{minipage}[c]{2.5cm} \[  U(1) Z_{30}^2\] \end{minipage}
\> \begin{minipage}[c]{6cm} \[ 9  \] \end{minipage} 
\> \begin{minipage}[c]{6cm} \[ 81 \]
\end{minipage} \\
\end{tabbing}

In this example none of the discrete anomaly matching conditions are
satisfied. Note that the global symmetries and charge
assignments in this theory are exactly equal to those in the
$SU(4)$ example at the beginning of this section because the
dimensions of the representations and the ratios of the Dynkin indices
are the same. However, the
values of the Dynkin
indices under the gauge groups are different (only their ratios are
the same), and
thus there is a different discrete symmetry in this theory.  Even
though the values of the discrete anomalies are exactly equal
in the two theories, the discrete anomaly matching conditions are very
different.

To close this section, we comment on the dual of 
non-supersymmetric QCD recently proposed by
Terning~\cite{John}. It has been suggested that $SU(3)$ with $F$
flavors of left- and right-handed quarks 
might have a dual in terms of a $G(F-6)$ group with $F$ flavors as well and
some composite baryons containing three quarks, and $G$ could be
$SU,Sp$ or $SO$. Since the electric theory is an $SU$ theory which
contains fundamentals, it does not have any interesting discrete
symmetries. If the dual gauge group is $SU(F-6)$ or $Sp(F-6)$ (for
even number of flavors) then the same statement holds for the dual
theory. However, if one assumes that $G=SO(F-6)$, then the dual theory
does have a $Z_{4F}$ non-trivial discrete symmetry, which can not be
mapped to any non-trivial discrete symmetry of the electric theory.
The lack of mapping of the discrete global symmetries  
makes the $SU(3)\leftrightarrow SO(F-6)$ duality much less
plausible, even though it can not be completely excluded due to the
potential presence of accidental symmetries in the strongly
interacting electric theory for $F<33/2$. However, for $F>33/2$, the
$SU(3)$ theory is infrared free and thus weakly coupled, and 
accidental symmetries cannot appear. Thus the 
$SU(3)\leftrightarrow SO(F-6)$ duality is certainly excluded for
$F>33/2$ and implausible for $F<33/2$.  We have to emphasize again,
however, that the  
$SU(3)\leftrightarrow SU(F-6)$ duality  is still a valid possibility
about which we have nothing new to say.

\subsection{Self-dual Theories}
The final set of examples we will consider are the $N=1$
supersymmetric self-dual examples based on certain exceptional groups
and $SO$ groups with spinors~\cite{Ramond,Distler,Karch,ourself}. 
(The self-dual theories of
Refs.~\cite{Sp,ourself} based on $SU$ and $Sp$ groups do not have any
non-trivial discrete symmetries and thus one can not gain new
information about them).

Let us consider, for example, the self-dual theory of Ref.~\cite{Ramond}
based on an $E_6$ gauge group. The conjectured electric and magnetic
theories are:
\begin{equation} \begin{array}{c|c|ccc}
& E_6 & SU(6) & U(1)_R & Z_{36} \\ \hline
Q & 27 & \Yfund & \frac{1}{3} & 1 \\ \hline \hline
q & 27 & \overline{\Yfund} & \frac{1}{3} & -1 \\
Z & 1 & \Ythrees & 1 & 3 \end{array} \end{equation}
with a superpotential in the magnetic theory $W=Zq^3$. The $Z_{36}$
charge of $Z$ has been chosen such that the mapping $Z\leftrightarrow
Q^3$ is obeyed, while that of $q$ such that the magnetic superpotential is
invariant under $Z_{36}$. Note  that one could add a multiple of $12$
to the $q$ charge. The Type I discrete anomalies are:

\begin{tabbing} \hspace*{2cm}
\= \hspace*{3cm} \= \hspace*{2.5cm} UV \hspace*{2cm} 
\= \hspace*{3cm}IR \hspace*{1cm} \\
\>  \begin{minipage}[c]{2.5cm} \[ SU(6)^2Z_{36} \] \end{minipage} 
\> \begin{minipage}[c]{6cm} \[ 27  \] \end{minipage} 
\> \begin{minipage}[c]{6cm}
\[ 81 \] \end{minipage} \\
\>  \begin{minipage}[c]{2.5cm} \[ Z_{36} ({\rm gravity})^2 \] \end{minipage} 
\> \begin{minipage}[c]{6cm} \[ 162 \] \end{minipage} 
\> \begin{minipage}[c]{6cm}
\[  6 \]
\end{minipage} \\
\end{tabbing}
Neither of these anomaly matching conditions is satisfied.  The
ambiguity of a multiple of $12$ in the $Z_{36}$ charge assignments for
$q$ does not help the anomaly matching either.

The failure
of anomaly matching could have been actually expected, since the $Z_{36}$
symmetries of the electric and the magnetic theories can not be mapped
to each other.  This is because the $E_6$ theory contains at least one
more independent flat direction corresponding to $Q^6$, which is
supposedly matched to $q^6$ of the magnetic theory.\footnote{A
  complete classification of gauge invariant polynomials is not
  known for exceptional groups.} This is however
impossible, since $Q^6$ carries $Z_{36}$ charge $6$, while $q^6$
charge $-6$. The difference of charges is $12$, and there is no way to
make up for this charge difference since there is no  other non-trivial
discrete symmetry in this theory. Thus we have to conclude that this
duality does not satisfy the mapping of global symmetries, unless we
assume that there are accidental $Z_{36}$ symmetries appearing both in
the electric and the magnetic theories (which is not impossible since 
both theories are strongly coupled). Therefore we conclude that the
lack of the matching 
of discrete global symmetries makes this duality much less plausible
even though this self-dual is not completely excluded. One possible
way to cure the lack of matching of the discrete $Z_{36}$ symmetries
in the above $E_6$ example is to modify the electric theory by adding
a tree-level superpotential $W=Q^6$, and regard this as a Kutasov-type
duality.\footnote{We thank Philippe Pouliot for pointing this out to
  us.} The magnetic superpotential then becomes $W=q^6+Zq^3$. The
additional superpotential terms explicitly break the $Z_{36}$ discrete
symmetry to $Z_6\subset SU(6)$ both in the electric and the magnetic
theories (and also break part of the $SU(6)$ global symmetries). 
This way the
constraints arising from the discrete symmetries are eliminated and
all the other consistency conditions for this duality are
satisfied.

One can show that the same statement holds for every self-dual theory
based on exceptional groups presented in
Refs.~\cite{Ramond,Distler,Karch}, that is without a tree-level
superpotential term the mapping of discrete symmetries is not
manifest, however after introducing tree-level superpotential terms
one obtains consistent Kutasov-type dualities.

There is another set of self-dual theories which have non-trivial
discrete symmetries: the $SO$ series of Ref.~\cite{ourself} and an
analogous $SO$ series of Ref.~\cite{Karch}. Let us, for example, examine
a self-dual theory from the $SO$ series of Ref.~\cite{ourself}.
Let us consider the $SO(12)$ theory with one spinor and eight
vectors. The dual pair is described by
\begin{equation} \begin{array}{c|c|cccc}
& SO(12) & SU(8) & U(1) & U(1)_R & Z_8 \\ \hline
S & 32 & 1 & 2 & \frac{1}{2} & 1 \\
V & \Yfund & \Yfund & -1 & 0 & 0 \\ \hline \hline
s & 32 & 1 & 2 & \frac{1}{2} & -1 \\
v & \Yfund & \Yfund & -1 & 0 & 0 \\
(S^2V^2) & 1 & \Yasymm & 2 & 1 & 2 \\
(S^2V^6) & 1 & \overline{\Yasymm} & -2 & 1 & 2 \end{array} \end{equation}
and a superpotential in the magnetic theory $W=(S^2V^2) s^2v^6+
(S^2V^6) s^2v^2$. One can see that it is possible to assign a $Z_8$ discrete
symmetry in the dual theory such that the superpotential is $Z_8$
invariant and the gauge singlets $(S^2V^2)$ and $(S^2V^6)$ have the
correct $Z_8$ charges. But this is not enough, since the mapping of
all other independent gauge invariants has to preserve $Z_8$ as
well. It turns out that this example does satisfy this additional
requirement. The reason is that the additional
independent gauge invariants involve only the fourth power of the
spinor $S$, and thus the $Z_8$ charge of such operators will be $\pm 4$
in the electric and the magnetic theory.  Discrete anomalies match
between the electric and magnetic theories almost trivially because of
the high multiplicity of the fields (32 for the spinor).  

One can also check
that the other examples in the $SO$ series in 
Ref.~\cite{ourself} do have the correct mapping of the discrete
symmetries once the maximal number of gauge singlet mesons are
included into the magnetic degrees of freedom. 
One way to see this is that most of the other self-dual
models of the $SO$ series can be derived from the above $SO(12)$
example by giving expectation values to vectors. The other way to see
it is to note that the discrete symmetry is $Z_8$ in every case, and
the gauge invariants contain only two or four powers of the spinor
field. If one includes all gauge invariants containing two powers of
spinors as elementary fields in the dual theory, the remaining operators
with four powers of spinors can be matched by the similar construction
as above. However, the self-duals in the $SO$ series of
Ref.~\cite{ourself} where not all of the invariants quadratic in spinors are
included as elementary fields in the magnetic theory cannot have the
required mapping of discrete symmetries. Thus the requirement of
discrete anomaly matching favors a single dual rather than multiple
duals. The remaining multiple self-duals can be interpreted only as
Kutasov-type dualities after adding a tree-level superpotential term
$S^4$ to the electric theory. 

Similarly, one can show that the $SO$ series of Ref.~\cite{Karch}
satisfy the discrete anomaly matching. The reason is that the
highest theory based on the $SO(14)$ group does have the correct
mapping of discrete symmetries, and all other examples can be derived
from this by giving an expectation value to one vector.
To see this, let us investigate
the $SO(14)$ example of~\cite{Karch}. The field content of the
electric and the dual magnetic theories is given by:
\begin{equation} \begin{array}{c|c|ccccc}
& SO(14) & SU(6) & U(1) & U(1)_R & Z_{12} \\ \hline
S & 64 & 1 & 3 & \frac{1}{7} & 0 \\
V & \Yfund & \Yfund & -4 & \frac{1}{7} & 1 \\ \hline \hline
s & 64 & 1 & 3 & \frac{1}{7} & 0 \\
v & \Yfund & \Yfund & -4 & \frac{1}{7} & -5 \\
(S^2V^3) & 1 & \Ythreea & -6 & \frac{5}{7} & 3 \\
(S^6V^3) & 1 & \Ythreea & 6 & \frac{9}{7} & 3 \end{array} \end{equation}
with a superpotential in the magnetic theory $W=(S^2V^3)s^6v^3+
(S^6V^3)s^2v^3$. The $Z_{12}$ charges of the additional gauge invariants
$S^4Q^2$, $S^4Q^4$ and $S^8Q^4$ are also mapped between the
electric and the magnetic theories. The only non-trivial discrete
anomaly is the $SU(6)^2Z_{12}$ which is $14$ in the electric and $-70$
in the magnetic theories; the difference is a multiple of $12$. All
other anomalies are multiples of $12$ themselves. 

In summary, the self-dual theories based on exceptional groups do not
have the correct mapping of discrete symmetries between the electric
and the magnetic theories, and hence have a much weaker foundation
than dualities where one does not have to rely on accidental
symmetries. However, they can have consistent interpretation as
Kutasov-type dualities, once additional terms are included in the superpotential.
The self-dual $SO$ series do have the correct mapping of
discrete symmetries and satisfy the discrete anomaly matching
conditions, once the maximal number of meson fields is included in the
magnetic theory. 

\section{Conclusions\label{sec:concl}}
\setcounter{equation}{0}
\setcounter{footnote}{0}

We have shown that any conjectured low-energy bound state spectrum 
has to satisfy anomaly matching conditions for the discrete global
symmetries. There are two types of discrete anomalies. Type I anomaly
matching conditions ($G_F^2Z_N$ and $Z_N({\rm gravity})^2$) have to be
satisfied
regardless of assumptions on the massive bound states. Type II
constraints have to be satisfied except if there are fractionally
charged massive states. 
We have given two separate arguments for discrete
anomaly matching. The argument based on instantons is valid only for
Type I anomalies, but it does not not involve any subtleties
concerning charge fractionalization. The argument based on spurions 
is valid for all anomalies, but the issues of 
charge fractionalization have to be taken into account.

We have tested several conjectured low-energy
solutions using discrete anomaly matching. All the results by Seiberg
on $N=1$ supersymmetric gauge theories satisfy these conditions,
which in some cases are extremely non-trivial. However, certain
solutions do not satisfy the discrete anomaly matching
conditions. These include an explicit realization of a 
chirally symmetric phase of $N=1$ pure
Yang-Mills theories based on the Veneziano--Yankielowicz Lagrangian,
several non-supersymmetric confining theories with large
representations, and 
some self-dual $N=1$ supersymmetric theories based on exceptional
gauge groups. These theories should be considered excluded or at least
highly implausible.

\section*{Acknowledgements}
We are grateful to Nima Arkani-Hamed,
Jan de Boer, Bob Cahn, Kentaro Hori, Shamit Kachru, Markus Luty, Hirosi 
Ooguri, Yaron Oz, Philippe Pouliot, Martin Schmaltz and 
Bruno Zumino for useful discussions, and to Ken Intriligator, John
March-Russell and Witold Skiba for comments on the manuscript. 
This work was 
supported in part by the U.S. Department of Energy under Contracts 
DE-AC03-76SF00098 and in part by the National Science Foundation under 
grant PHY-95-14797. C.C. is a research fellow of the Miller Institute
for Basic Research in Science. H.M. is an Alfred P. Sloan Foundation
fellow. 

\appendix
\renewcommand{\thesection}{Appendix \Alph{section}}
\renewcommand{\theequation}{\Alph{section}.\arabic{equation}}

\section{Outer Automorphism, Charge Conjugation, and Color Parity 
in $SO(N)$ Groups\label{app:charge}}
\setcounter{equation}{0}
\setcounter{footnote}{0}

We have seen in Section~\ref{sec:discrete} that
 $SO(2n)$ groups have a non-trivial
$Z_2$ outer automorphism. The definition we gave for this automorphism
is a parity-like transformation ${\cal P}$ (color-parity)
in the internal $2n$ dimensional space which flips the sign 
of one particular color. 
In this appendix, we show that color-parity defines an 
outer automorphism for all $SO(2n)$ groups.  The 
usual charge conjugation $T^a\to -T^{a*}$ is equivalent to 
color-parity for $SO(4k+2)$ groups while it is trivial up to a gauge 
transformation for $SO(4k)$ groups.

We first explain our notation for $SO(2n)$ groups~\cite{WZ}. The
$2^n$ by $2^n$ gamma matrices $\gamma_i$ form a Clifford algebra
\begin{equation} \{\gamma_i,\gamma_j\} =2\delta_{ij}, 
\qquad i,j=1,2,\ldots ,2n. \label{eq:Clifford}
\end{equation}
These gamma matrices can be constructed by iteration starting with
$\gamma_1=\tau_1$ and $\gamma_2=\tau_2$ for $n=1$ and taking tensor
products with $\tau_3$ (the $\tau$'s are the Pauli matrices).
The explicit form of the general $\gamma$ matrices is then
\begin{eqnarray}
\gamma_{2k-1}&=&\overbrace{\tau_3 \otimes \tau_3 \otimes \ldots 
\otimes \tau_3}^{n-k} \otimes \tau_1 
\otimes \overbrace{1 \otimes 1 \otimes \ldots \otimes 1}^{k-1},
  \\
\gamma_{2k}&=&{\tau_3 \otimes \tau_3 \otimes \ldots \otimes
  \tau_3} \otimes
\tau_2 \otimes 
{1 \otimes 1 \otimes \ldots \otimes 1}
\end{eqnarray}
with $1$ appearing $k-1$ times in the product, $\tau_3$ appearing
$n-k$ times in the product and $\tau_2$ or $\tau_1$ appearing once.
This way we can see that $\gamma_k$ is antisymmetric for even $k$ 
while $\gamma_k$ is symmetric for odd $k$. The spinor representation
of $SO(2n)$ is defined as the object transforming as 
\begin{equation} 
\psi \to e^{i\omega_{ij}\sigma_{ij}/4}\psi, \qquad 
\sigma_{ij}= \frac{i}{2}
[\gamma_i, \gamma_j].
\end{equation}
The analog of $\gamma_5$ is $\gamma_{2n+1}$ which is defined by
\begin{equation} \gamma_{2n+1}=(-i)^n \gamma_1\gamma_2\ldots \gamma_{2n}=
\tau_3\otimes \tau_3 \otimes \ldots \otimes \tau_3,\end{equation}
which anticommutes with all $\gamma_i$'s. Thus the spinors
$\frac{1}{2}(1+\gamma_{2n+1}) \psi$ and  $\frac{1}{2}(1-\gamma_{2n+1})
\psi$ transform
separately and there are two inequivalent spinor representations for
$SO(2n)$. 

The usual definition of the charge conjugation matrix $C$ in field
theory is that $\psi_{1}^{T} C\psi_{2}$ is invariant under $SO(2n)$
transformations. This implies that 
\begin{equation}
\label{socharge}
 C^{-1} \sigma^T_{ij}C=-\sigma_{ij},
\end{equation}
which exactly coincides with the definition $T^a\to -T^{a*}$ (\ref{cc})
in the
spinor basis. Note that this implies that 
\begin{equation} 
  C^{-1} [\gamma_i^T,\gamma_j^T]C=[\gamma_i,\gamma_j],
\end{equation}
for the $\gamma$ matrices, which is satisfied if
\begin{equation} C^{-1}\gamma_i^TC=\pm \gamma_i.\end{equation}
This equation can be satisfied either with
\begin{equation} C_1= \gamma_1\gamma_3\ldots \gamma_{2n-1}
\label{eq:C1}\end{equation}
or with 
\begin{equation} C_2= \gamma_2\gamma_4\ldots \gamma_{2n}.
\label{eq:C2}\end{equation}
These are both good definitions of charge conjugation in the sense of 
Eq.~(\ref{socharge}), but their symmetry properties are
different.\footnote{Usually, the matrix which satisfies $C^{-1} 
\gamma_{i}^{T} C = - \gamma_{i}$ is referred to as the charge 
conjugation matrix, while the other matrix which satisfies $T^{-1} 
\gamma_{i}^{T} T = \gamma_{i}$ as the time reversal matrix.}
For $C_1$:
\begin{equation}  C_1^{-1}\gamma_i^TC_1= (-1)^{n-1}\gamma_i.\end{equation}
To see this relation let us first assume that the subscript 
$i$ is even. Then $\gamma_i^T=-\gamma_i$, and one has to
anticommute $\gamma_i$ with $n$ different $\gamma$ matrices, thus the
sign $(-1)^{n-1}$. Similarly, $\gamma_i$ is
symmetric if $i$ is odd, but one has to perform only $n-1$ exchanges,
thus the above 
formula follows for $i$ even or odd. Similarly, one can show that for 
$C_2$:
\begin{equation}  C_2^{-1}\gamma_i^TC_2= (-1)^{n}\gamma_i.\end{equation}
From the above explicit construction of the $C$'s in terms of $\gamma$
matrices and from the iterative construction for the $\gamma$ matrices,
we obtain that 
\begin{eqnarray}
&& C_1= i\tau_2\otimes \tau_1\otimes i\tau_2\otimes \tau_1\otimes \ldots 
\otimes i\tau_2\otimes \tau_1\nonumber \\
&& C_2= -i\tau_1\otimes \tau_2\otimes -i\tau_1\otimes \tau_2\otimes \ldots 
\otimes -i \tau_1\otimes \tau_2 \end{eqnarray}
for $SO(4k)$, with $k$ $\tau_1$'s and $\tau_2$'s appearing both in
$C_1$ and in $C_2$, while for $SO(4k+2)$ 
\begin{eqnarray}
&& C_1= \tau_1\otimes i\tau_2\otimes \tau_1\otimes i\tau_2\otimes  
\tau_1\otimes \ldots 
\otimes i\tau_2\otimes \tau_1\nonumber \\
&& C_2= \tau_2\otimes -i\tau_1\otimes \tau_2\otimes -i\tau_1\otimes
\tau_2\otimes  \ldots 
\otimes -i\tau_1\otimes \tau_2 \end{eqnarray} 
with $k+1$ factors of $\tau_1$'s and $k$ factors of $\tau_2$'s 
appearing in $C_1$ while in $C_2$ there are  
$k+1$ factors of $\tau_2$'s and $k$ factors of $\tau_1$'s. 

Armed with this knowledge we can now proceed to show that 
the effect of charge conjugation is just equivalent to an
$SO(4k)$ gauge transformation for $SO(4k)$ gauge groups. 
For this all we need is that $C\in SO(4k)$ for the above
choice of the charge conjugation matrix $C$. Let
us choose, for example,
$C=C_1= \gamma_1\gamma_3\ldots \gamma_{4k-1}$. Since we are
considering $SO(4k)$ groups, there are even number of $\gamma$
matrices appearing in $C$, thus $C$ can also be written as
$C=(\gamma_1\gamma_3)(\gamma_5\gamma_7)\ldots
(\gamma_{4k-3}\gamma_{4k-1})$. However, the products
$\gamma_i\gamma_{i+2}$ are all $SO(4k)$ elements. This can be seen by
considering the $SO(4k)$ element 
\begin{equation}
O(i,j)=e^{-i\frac{\pi}{2} \sigma_{ij}}=\cos \frac{\pi}{2}-i \sigma_{ij}
\sin \frac{\pi}{2}=-i\frac{i}{2}[\gamma_i,\gamma_{j}]
=\gamma_i\gamma_{j} \qquad (i\neq j)
\end{equation}
due to the anticommutation relations of the $\gamma$'s. Thus for
$SO(4k)$ groups, $C = O(1,3) O(5,7)$ $\cdots O(4k-3,4k-1)$ 
is an element of the gauge group, and does not
act as an outer automorphism on the Lie algebra (it is not an
additional discrete symmetry of the theory). This proof shows at the same
time that the spinors for $SO(4k)$ are self conjugates (real or
pseudo-real), since
charge conjugation is equivalent to a gauge transformation and 
does not interchange representations. 

The situation is very different for $SO(4k+2)$, since there $C$
contains odd number of $\gamma$ matrices and therefore can not be a
gauge transformation. This can be most easily seen by the fact that
$C$ anticommutes with $\gamma_{2n+1}$ and thus interchanges the two
kinds of spinor representations. The two kinds of 
spinors of $SO(4k+2)$, therefore, are charge conjugates of each other. 

What remains to be seen is that the color-parity transformation ${\cal
  P}$ which 
we employed to define the outer automorphism for $SO(2n)$ coincides 
with the above charge conjugation up to a gauge transformation for 
$SO(4k+2)$, and is also a non-trivial automorphism for $SO(4k)$.  We 
have defined the automorphism ${\cal P}$ of $SO(2n)$ as a parity-like 
transformation in the internal $2n$ dimensional space, which flips the 
sign of a particular color ({\it e.g.}\/ 1).  This means that the transformed 
spinor $\psi'$ has to transform as 
\begin{equation}
\psi'\to e^{i\omega_{ij}'\sigma_{ij}/4}\psi'
\end{equation} 
under an $SO(2n)$ transformation, 
where $\omega_{1i}'=-\omega_{1i}$ and $\omega_{ij}' = \omega_{ij}$ for 
$i,j\neq 1$.  Since $\gamma_1\gamma_i\gamma_1=-\gamma_i$ if $i\neq 1$, 
the spinor constructed by $\psi'=i\gamma_1 \gamma_{2n+1}\psi$ will 
transform exactly the right way.  Thus we conclude that parity-like 
transformation which changes the sign of the $1$ direction is 
implemented on the spinors by multiplication by $i\gamma_1\gamma_{2k+1}$.  
Since 
$\gamma_1$ always anticommutes with $\gamma_{2n+1}$, it connects the 
two different spinor representations characterized by 
$\gamma_{2n+1}=\pm 1$, and thus cannot be an inner 
automorphism.  Therefore this transformation is always a good 
definition for the outer automorphism of $SO(2n)$.  Note that a 
similar definition of automorphism for $SO(2n+1)$ groups is gauge 
equivalent to the overall sign flip of the whole vector
in the $SO(2n+1)$ group, and is of flavor-type.  The transformation of
the spinors under this flavor-type symmetry depends on the models.

Finally, we show that the two definitions for
the Lie algebra automorphisms coincide for $SO(4k+2)$ up to gauge
transformations. We have 
seen that the definition using the parity-like transformation acts 
as a multiplication by $i\gamma_1\gamma_{4k+3}=(-1)^{k}\gamma_{2} 
\gamma_{3} \cdots \gamma_{4k+2}$ on the spinor, while the charge
conjugation acts like multiplication by $C$. The important point is
that $i\gamma_{1}\gamma_{4k+3}=(\gamma_{3}\gamma_{5})\cdots 
(\gamma_{4k-1}\gamma_{4k+1})C_{2}$.  A pair of
$\gamma$ matrices can be thought of as an $SO(4k+2)$ transformation as
we have shown above. Thus $C_{2}$ is nothing but a product of
$SO(4k+2)$ transformation matrices multiplied by the $i\gamma_1\gamma_{4k+3}$
matrix, and is hence equivalent to the
color-parity transformation up to a gauge transformation.  The other 
charge conjugation matrix $C_{1}=i\gamma_{4k+3}C_{2}^{-1}$ is also 
gauge equivalent because $i\gamma_{4k+3}=(-1)^{k}O(1,2) \cdots 
O(4k+1,4k+2)$.

Let us add brief comments on $SO(2n+1)$ groups.  We add 
$\gamma_{2n+1}$ to the set of $\gamma_{i}$-matrices to represent the 
Clifford algebra (\ref{eq:Clifford}) for $i=1, \cdots, 2n+1$.  The 
matrix $C_{2}$ (\ref{eq:C2}) satisfies $C_{2}^{-1} \gamma_{i}^{T} 
C_{2} = (-1)^{n} \gamma_{i}$ including $i=2n+1$.  There is, however, 
no consistent definition of the color-parity for $SO(2n+1)$ spinors.  
On the other hand, color-parity can be defined on the vectors as the sign
flip of the first color. Together with a gauge 
transformation $O(2,3) O(4,5) \cdots O(2n,2n+1)$, however, this color-parity
flips the sign of a 
vector as a whole, and hence is of flavor-type ({\it i.e.}\/, 
commutes with the gauge group).  This is because there is no outer 
automorphism for $SO(2n+1)$ groups.  Therefore, the only possible 
discrete symmetries in $SO(2n+1)$ gauge theories are flavor-type 
symmetries.

\section{Centers of Simple Groups\label{app:centers}}
\setcounter{equation}{0}
\setcounter{footnote}{0}

We have seen in Section~\ref{sec:discrete} that the correct 
identification of the additional discrete symmetries requires the 
knowledge of the centers of the continuous global symmetries.  One can 
avoid unnecessary checks of anomaly matching when a discrete symmetry 
is a part of the continuous ones.  The centers of semi-simple Lie 
groups have been classified (see e.g.~\cite{auto}) 
and in the following we give a complete list of them:
\begin{eqnarray*}
& SU(N): & Z_N  \\
& Sp(2N): & Z_2  \\
& SO(2N+1): & Z_2  \\
& SO(4N): & Z_2\times Z_2  \\
& SO(4N+2): & Z_4  \\
& E_6: & Z_3  \\
& E_7: & Z_2 \end{eqnarray*} 
The other semi-simple groups do not have a non-trivial center. 

Let us
give what the actions of the centers are. For $SU(N)$, the center is
the $Z_N$ phase rotation of the fundamental representation. The
$Z_2$ center of $Sp(2N)$ is the sign flip of the fundamental of
$Sp(2N)$.  For
$SO(2N+1)$ the center is the $2\pi$ rotation in the $SO(2N+1)$
gauge group which flips the sign of the spinor representation. 

The case of $SO(2n)$ groups is more complicated. The center is
different for $SO(4k)$ and $SO(4k+2)$, which has to do with the
different definition of the analog of $\gamma_5$,
$\gamma_{2n+1}=(-i)^n\gamma_1\gamma_2\ldots \gamma_{2n}$. 
For $SO(4k)$ groups ($n=2k$), $\gamma_{4k+1}=
(-1)^k \gamma_1\gamma_2 \ldots \gamma_{4k}$. Note that there is no $i$
in the definition of $\gamma_{4k+1}$. As shown in
\ref{app:charge}, a
product of two $\gamma$-matrices $\gamma_i\gamma_j$ is always an $SO$
group element $O(i,j)$, which is just a 180 degree rotation in the
$i-j$ plane. We know however, how $\gamma_{4k+1}$ acts on the spinors:
$\gamma_{4k+1}S_1=S_1$, $\gamma_{4k+1}S_2=-S_2$. Since 
$\gamma_{4k+1}=(-1)^k O(1,2)O(3,4)  \ldots   O(4k-1,4k)$, we conclude that
the $SO(4k)$ group element $g=O(1,2)O(3,4)  \ldots   O(4k-1,4k)$ acts
as the above $Z_2$ transformation on the spinors.  Note that this
$SO(4k)$ element flips the overall sign of the vector.  There is a separate
$Z_2$ transformation: $2\pi$ $SO(4k)$ rotation that flips
the signs of both spinors, $S_1\to -S_1$, $S_2\to -S_2$.  Note 
that the vector of 
$SO(4k)$ switches sign under the $g=O(1,2)O(3,4)\ldots O(4k-1,4k)$
$SO(4k)$ transformation, but not under the $2\pi$ $SO(4k)$
rotation.  We can
combine the $Z_2$ of $2\pi$ rotation and the other $Z_2$ 
generated by $g$ to obtain the $Z_2\times Z_2$ center of $SO(4k)$ as
sign flips of any two of the spinors $S_1$, $S_2$ and the vector.

The case of $SO(4k+2)$ groups differs from the $SO(4k)$ because 
$\gamma_{4k+3}=i(-1)^k$ $O(1,2)$ $O(3,4)$ $\ldots O(4k+1,4k+2)$. Thus
the effect 
of $g'=O(1,2)O(3,4)\ldots O(4k+1,4k+2)$ on the spinors is $S_1\to iS_1$,
$S_2\to -iS_2$, which forms a $Z_4$ group that is the center
of $SO(4k+2)$. The vector of $SO(4k+2)$ has charge two under the $Z_4$
center.

The center of $E_6$ is $Z_3$. We can find the action of the center on the
representations by considering the embedding of $SO(10)\times U(1)$
into $E_6$. Under this subgroup $27\to 16_1+10_{-2}+1_4$ where the
lower indices are the $U(1)$ charges. Let us consider the $Z_3$
subgroup of $U(1)$ under which 
\begin{eqnarray}
&& 16\to e^{2\pi i \frac{n}{3}}16, \nonumber \\
&& 10\to e^{2\pi i \frac{-2n}{3}}10= e^{2\pi i \frac{n}{3}}10,
\qquad\qquad n=0,1,2 \nonumber \\
&& 1\to e^{2\pi i \frac{4n}{3}}10= e^{2\pi i \frac{n}{3}}1 .
\end{eqnarray}
This $Z_3$ acts uniformly on $16,10$ and $1$ and is a $Z_3$
phase rotation of the fundamental $27$ of $E_6$. Therefore this $Z_3$ is
the center of $E_6$.

The center of $E_7$ is $Z_2$. To identify the action of this $Z_2$, we
note that $E_7$ has an $SU(8)$ subgroup under which the fundamental
$56$ decomposes as $56\to 28+\overline{28}$, where $28$ is the rank two
antisymmetric tensor of $SU(8)$. The center of $SU(8)$ is $Z_8$, whose
action on $28$ is effectively a $Z_4$. However, $28$ and
$\overline{28}$ transform
with the opposite phase under $Z_4$, thus only the $Z_2$ subgroup of
$Z_4$ acts uniformly on $28$ and $\overline{28}$. Therefore the $Z_2$ center of
$E_7$ is the sign flip of the fundamental $56$ representation.


\begin{thebibliography}{99}

\bibitem{tHooft} 
G. 't Hooft, in ``Recent Developments in Gauge
  Theories'' eds. G. 't Hooft et. al. (Plenum Press, New York, 1980), 135.

\bibitem{Seiberg} 
N. Seiberg, \PRD{49}{6857}{1994}, hep-th/9402044;
\NPB{435}{129}{1995}, hep-th/9411149; K. Intriligator, R. Leigh and
N. Seiberg, \PRD{50}{1092}{1994}. 

\bibitem{IntrSeib} 
K. Intriligator and N. Seiberg, \NPB{444}{125}{1995}, hep-th/9503179.

\bibitem{IntrPoul}
K. Intriligator and P. Pouliot, \PLB{353}{471}{1995}, hep-th/9505006.

\bibitem{Kutasov} 
D. Kutasov, \PLB{351}{230}{1995}, hep-th/9503086; D. Kutasov and
A. Schwimmer, \PLB{354}{315}{1995}, hep-th/9505004; D. Kutasov,
A. Schwimmer and N. Seiberg, \NPB{459}{455}{1996}.

\bibitem{otherKutasov} 
K. Intriligator, R. Leigh and M. Strassler,
  \NPB{456}{567}{1995}, hep-th/9506148; J. Brodie and M. Strassler,
  hep-th/9611197. 

\bibitem{G2}
I. Pesando, \MPLA{10}{1871}{1995}, hep-th/9506139; S. Giddings and
J. Pierre, \PRD{52}{6065}{1995}, hep-th/9506196.

\bibitem{SOduals}
P. Pouliot, \PLB{359}{108}{1995}, hep-th/9507018; P. Pouliot and
M. Strassler, \PLB{370}{76}{1996}, hep-th/9510228;
\PLB{375}{175}{1996}, hep-th/9602031; M. Berkooz, P. Cho, P. Kraus and
M. Strassler, hep-th/9705003.

\bibitem{deconf}
M. Berkooz, \NPB{452}{513}{1995}, hep-th/9505067; M. Luty, M. Schmaltz
and J. Terning, \PRD{54}{7815}{1996}, hep-th/9603034.


\bibitem{SUantisymm}
H. Murayama, \PLB{355}{187}{1995}, hep-th/9505082; E. Poppitz and
S. Trivedi, \PLB{365}{125}{1996}, hep-th/9507169; 
P. Pouliot, \PLB{367}{151}{1996}, hep-th/9510148.

\bibitem{sconf} 
C. Cs\'aki, M. Schmaltz and W. Skiba,
  \PRL{78}{799}{1997}, hep-th/9610139; \PRD{55}{7840}{1997}, hep-th/9612207.

\bibitem{DM}
G. Dotti and A. Manohar, hep-th/9706075; hep-th/9710024; G. Dotti,
hep-th/9709161. 


\bibitem{GN}
B. Grinstein and D. Nolte, hep-th/9710001.

\bibitem{Shifman} 
A. Kovner and M. Shifman, \PRD{56}{2396}{1997}, hep-th/9702174.

\bibitem{VY} 
G. Veneziano and S. Yankielowicz, \PLB{113}{231}{1982};
T. Taylor, G. Veneziano and S. Yankielowicz, \NPB{218}{493}{1983}.

\bibitem{Albright} 
C. Albright, \PRD{24}{1969}{1981}.

\bibitem{othernonsusy}
T. Banks, S. Yankielowicz and A. Schwimmer, \PLB{96}{67}{1980}; 
S. Dimopoulos, S. Raby and L. Susskind, \NPB{173}{208}{1980};
\NPB{169}{373}{1980}.

\bibitem{Ramond} 
P. Ramond, \PLB{390}{179}{1997}, hep-th/9608077.

\bibitem{Distler} 
J. Distler and A. Karch, hep-th/9611088.

\bibitem{Karch} 
A. Karch, \PLB{405}{280}{1997}, hep-th/9702179.

\bibitem{tHooft2} 
G. 't Hooft, \PRL{37}{8}{1976}; \PRD{14}{3432}{1976}.


\bibitem{auto}
N. Bourbaki, ``Lie Groups and Lie Algebras,'' (Hermann, Paris, 1975).

\bibitem{PTWW} 
J. Preskill, S. Trivedi, F. Wilczek and M. Wise,
  \NPB{363}{207}{1991}. 

\bibitem{IR} 
L. Ib\'a\~nez and G. Ross, \PLB{260}{291}{1991}; \NPB{368}{3}{1992};
L. Ib\'a\~nez, \NPB{398}{301}{1993}, hep-ph/9210211; 

\bibitem{BD} 
T. Banks and M. Dine, \PRD{45}{1424}{1992}, hep-th/9109045.

\bibitem{KW}
L. M. Krauss and F. Wilczek \PRL{62}{1221}{1989}; J. Preskill and 
L. M. Krauss, \NPB{341}{50}{1990}.

\bibitem{Rohlin}
V. A. Rohlin, {\sl Dok. Akad. Nauk. USSR}\/, {\bf
  84}, 221 (1952).

\bibitem{EGH} 
T. Eguchi, P. B. Gilkey, and A. J. Hanson,
  {\sl Phys. Rep.}\/ {\bf 66}, 213 (1980).


\bibitem{KS} K. Konishi and K. Shizuya, \NC{90A}{111}{1985}.


\bibitem{ISS}
K. Intriligator, N. Seiberg and S. Shenker, \PLB{342}{152}{1995}, 
hep-ph/9410203.
 
\bibitem{Intrdual} 
K. Intriligator, \NPB{448}{187}{1995}, hep-th/9505051.


\bibitem{ourself} 
C. Cs\'aki, M. Schmaltz, W. Skiba and J. Terning,
  \PRD{56}{1228}{1997}, hep-th/9701191.

\bibitem{John} J. Terning, hep-th/9706074.

\bibitem{Sp}
C. Cs\'aki, W. Skiba and M. Schmaltz, \NPB{487}{128}{1997}, hep-th/9607210.

\bibitem{pureYM}
V. Novikov, M. Shifman, A. Vainshtein and V. Zakharov,
\NPB{229}{407}{1983}; M. Shifman and A. Vainshtein, \NPB{296}{445}{1988}.


\bibitem{WZ} 
F. Wilczek and A. Zee, \PRD{25}{553}{1982}.

\end{thebibliography}
\end{document}